\documentclass[a4paper,notitlepage]{article}
\usepackage{amssymb}

\usepackage{latexsym}
\usepackage{amsmath}
\usepackage{graphicx}

\newtheorem{theorem}{Theorem}

\newtheorem{corollary}[theorem]{Corollary}

\newtheorem{definition}[theorem]{Definition}

\newtheorem{lemma}[theorem]{Lemma}

\newtheorem{proposition}[theorem]{Proposition}
\newtheorem{remark}[theorem]{Remark}

\newenvironment{proof}[1][Proof]{\textbf{#1.} }{\ \rule{0.5em}{0.5em}}
\input{tcilatex}

\begin{document}

\title{Asymptotics of solutions in $nA+nB\rightarrow C$ reaction--diffusion systems}
\author{Guillaume van Baalen\thanks{%
Supported in part by the Fonds National Suisse.} \\
D\'{e}partement de Physique\\
Th\'{e}orique\\
Universit\'{e} de Gen\`{e}ve\\
Switzerland\\
vanbaal4@kalymnos.unige.ch \and Alain Schenkel\thanks{%
Supported by the Fonds National Suisse and DOE grant SCUOCB341495.} \\
Helsinki Institute of Physics\\
University of Helsinki\\
Finland\\
alain.schenkel@helsinki.fi \and Peter Wittwer\thanks{%
Supported in part by the Fonds National Suisse.} \\
D\'{e}partement de Physique\\
Th\'{e}orique\\
Universit\'{e} de Gen\`{e}ve\\
Switzerland\\
wittwer@ibm.unige.ch}
\maketitle

\begin{abstract}
We analyze the long time behavior of initial value problems that model a
process where particles of type $A$ and $B$ diffuse in some substratum and
react according to $nA+nB\rightarrow C.$ The case $n=1$ has been studied
before; it presents nontrivial behavior on the reactive scale only. In this
paper we discuss in detail the cases $n>3,$ and prove that they show
nontrivial behavior on the reactive and the diffusive length scale.\newpage
\end{abstract}

\tableofcontents

\section{Introduction and main results}

We consider, for arbitrary but fixed $n\in\mathbf{N,}$ $n\geq1,$ the
reaction--diffusion problem 
\begin{align}
a_{t} & =a_{xx}-\frac{1}{2}(4ab)^{n}~,  \label{system1} \\
b_{t} & =b_{xx}-\frac{1}{2}(4ab)^{n}~,  \label{system2}
\end{align}
for $x\in\mathbf{R},$ $t\geq\tau\geq0,$ with initial conditions $a(x,\tau
)=a_{0}(x),$ $b(x,\tau)=b_{0}(x),$ satisfying 
\begin{align}
\lim_{x\rightarrow-\infty}a_{0}(x) & =1~,  \notag \\
\lim_{x\rightarrow+\infty}b_{0}(x) & =1~,  \label{lim002} \\
\lim_{x\rightarrow+\infty}a_{0}(x) & =\lim_{x\rightarrow-\infty}b_{0}(x)=0~.
\notag
\end{align}
The choice of the initial time $t=\tau,$ and a class of initial conditions $%
a_{0},$ $b_{0}$ will be described later on, but for the purpose of this
introduction it is useful to have in mind the ``natural'' case: $\tau=0,$ $%
a_{0}(x)=1$ for $x<0,$ $a_{0}(x)=0$ for $x>0,$ and $b_{0}(x)=1$ for $x>0,$ $%
b_{0}(x)=0$ for $x<0.$

This initial value problem models the time evolution of a chemical system of
two (initially separated) substances $A$ and $B,$ that diffuse in some
substratum and react according to $nA+nB\rightarrow C,$ with a substance $C$
that is supposed not to participate in the reaction anymore. The model is a
mean--field description of such a situation where the functions $a$ and $b$
represent the densities of the substances $A$ and $B.$ For more details see 
\cite{Droz}.

Equations (\ref{system1}) and (\ref{system2}) are best studied in terms of
the sum 
\begin{equation}
v=a+b~,  \label{defv}
\end{equation}
and the difference 
\begin{equation}
u=a-b~,  \label{defu}
\end{equation}
which satisfy the equations 
\begin{align}
u_{t} & =u_{xx}~,  \label{equ} \\
v_{t} & =v_{xx}-(v^{2}-u^{2})^{n}~,  \label{eqv}
\end{align}
with initial conditions $v_{0}$ and $u_{0}$ (at time $t=\tau)$ that satisfy 
\begin{align}
\lim_{x\rightarrow-\infty}u_{0}(x) & =1~,  \notag \\
\lim_{x\rightarrow+\infty}u_{0}(x) & =-1~,  \label{limitu}
\end{align}
and 
\begin{equation}
\lim_{x\rightarrow\pm\infty}v_{0}(x)=1~.  \label{limitv}
\end{equation}
For initial conditions $a_{0},$ $b_{0}$ with 
\begin{equation}
a_{0}(x)=b_{0}(-x)~,  \label{sym001}
\end{equation}
the functions $v_{0}$ and $u_{0}$ are even and odd, respectively, and the
equations (\ref{equ}), (\ref{eqv}) preserve this symmetry. Furthermore, for
the special initial condition 
\begin{equation}
u(x,\tau)=-\mu_{1}(x/\sqrt{\tau})~,  \label{initsym}
\end{equation}
with $\mu_{1}$ defined by the equation 
\begin{equation}
\mu_{1}(y)=\mathrm{erf}(\frac{y}{2})\equiv\frac{2}{\sqrt{\pi}}%
\int_{0}^{y/2}e^{-\sigma^{2}}d\sigma~,  \label{defmu}
\end{equation}
equation (\ref{equ}) has the explicit solution 
\begin{equation}
u(x,t)=-\mu_{1}(x/\sqrt{t})~.  \label{uxt}
\end{equation}
We note that the initial condition (\ref{initsym}) for $u$ (at time $t=\tau$%
) is simply the solution of equation (\ref{equ}) with the ``natural''
initial condition, $u(x,0)=1$ for $x<0,$ $u(x,0)=-1$ for $x>0,$ evaluated at 
$t=\tau.$ To keep this paper as simple as possible we now restrict the
discussion to this case, i.e., we consider from now on equation (\ref{eqv})
with initial conditions satisfying (\ref{limitv}), and $u$ given by (\ref
{uxt}). We note, however, that more general (asymmetric) initial conditions
for $u$ could be treated as well. This would lead to corrections to $u$ of
the order $\mathcal{O}(1/t),$ and such corrections do not change in any way
the discussion of the equation for $v$ that follows.

The reaction--diffusion problems considered here develop, in addition to the
built--in diffusive length scale $\mathcal{O}(\sqrt{t}),$ an additional
shorter length scale, on which the reaction takes place. The function $F,$%
\begin{equation}
F=\frac{1}{2}(4ab)^{n}\equiv \frac{1}{2}(v^{2}-u^{2})^{n}~,
\label{reaction001}
\end{equation}
is called the reaction term or reaction front, and we are interested in
describing the asymptotic behavior of the function $F$ for large times. The
knowledge of this behavior is useful, since it appears to be universal, in
the sense that it is largely independent of the choice of the initial
conditions and of the details of the model under consideration. As mentioned
above, if $v_{0}$ is an even function, then $v$ and as a consequence $F$ are
even functions of $x.$ We will see that the critical point of $F$ at $x=0$
is a maximum, and that $F$ decays (rapidly) for large $x.$

Before proceeding any further we note that the factor of $4^{n-1/2}$ in (\ref
{system1}), (\ref{system2}) and (\ref{reaction001}) is just a normalization,
and has been chosen for convenience to make the equation (\ref{eqv}) for $v$
look simple. In fact, any system of the form 
\begin{align*}
a_{t} & =D_{a}a_{xx}-k_{a}(ab)^{n}~, \\
b_{t} & =D_{b}b_{xx}-k_{b}(ab)^{n}~,
\end{align*}
with positive $D_{a},$ $D_{b},$ $k_{a},$ and $k_{b},$ and with initial
conditions such that 
\begin{align*}
\lim_{x\rightarrow-\infty}a(x,0) & =a_{\infty}>0~, \\
\lim_{x\rightarrow\infty}b(x,0) & =b_{\infty}>0~, \\
\lim_{x\rightarrow+\infty}a(x,0) & =\lim_{x\rightarrow-\infty}b(x,0)=0~,
\end{align*}
can be reduced, by scaling space and time and the amplitudes, to the problem

\begin{align*}
a_{t} & =a_{xx}-\frac{1}{2}(4ab)^{n}~, \\
b_{t} & =Db_{xx}-\frac{1}{2}(4ab)^{n}~,
\end{align*}
with $D>0,$ and with initial conditions such that 
\begin{align*}
\lim_{x\rightarrow-\infty}a(x,0) & =1~, \\
\lim_{x\rightarrow\infty}b(x,0) & =\beta>0~, \\
\lim_{x\rightarrow+\infty}a(x,0) & =\lim_{x\rightarrow-\infty}b(x,0)=0~.
\end{align*}
In this paper we have limited the discussion to the case $\beta=1$ and $D=1. 
$ The case $\beta\neq1$ leads to a moving reaction front. A change of
coordinates to a moving frame complicates the analysis, but the problem
could still be treated with the methods presented here. Choosing $D=1$ makes
the equations mathematically simpler. As a consequence, as we have seen, the
two equations for $a$ and $b$ can be reduced to just one equation for the
sum $v=a+b,$ since the equation for the difference $u=a-b$ can be solved
explicitly. Even though we do not expect the asymptotic behavior of the
solution to change in any relevant way if $D\neq1$, the strategy of proof
would have to be changed considerably, since the equations can not be
decoupled anymore in that case.

Before we state our results, we briefly discuss the expected dependence of
the results on the parameter $n.$

The case $n=1$ has been studied in detail in \cite{Wittwer}, where it is
proved that in this case the reaction term (\ref{reaction001}) satisfies,
for all $z\in\mathbf{R},$ 
\begin{equation*}
\lim_{t\rightarrow\infty}t^{2\gamma}F(t^{\alpha}z,t)=\rho(\left| z\right| )~,
\end{equation*}
where $\alpha=1/6,$ and $\gamma=1/3,$ and where $\rho\colon\mathbf{R}_{+}%
\mathbf{\rightarrow R}_{+}$ is a smooth function that decays like $\exp(-%
\mathrm{const.}z^{3/2})$ for large values of $z.$ It follows furthermore
from the results in \cite{Wittwer} that the function $F$ is very small on
the diffusive scale in the sense that for $n=1,$ $y\neq0,$ and all $p\geq0, $%
\begin{equation}
\lim_{t\rightarrow\infty}t^{p}F(\sqrt{t}y,t)=0~.  \label{fdiffuse}
\end{equation}
The smallness of $F$ on the diffusive scale is easily understood by
realizing that, for $n=1$ and for positive values of $x$ on the diffusive
scale, i.e., for $x/\sqrt{t}>>1,$ equation (\ref{system1}) essentially
reduces to 
\begin{equation}
a_{t}=a_{xx}-\lambda a~,  \label{a001}
\end{equation}
with $\lambda>0.$ Therefore, the function $a$ decays exponentially fast to
zero on this scale, and similarly for $b$ for negative values of $x.$

For $n>1,$ however, equation (\ref{system1}) reduces, for $x/\sqrt{t}>>1,$
essentially to 
\begin{equation}
a_{t}=a_{xx}-\lambda a^{n}~,  \label{a002}
\end{equation}
with $\lambda>0.$ The solution of (\ref{a002}) has an asymptotic behavior
that is radically different from the solution of (\ref{a001}). In
particular, for $n=2$, the solution may even blow up in finite time if $a$
is not a positive function. Note that, for $n$ odd, the nonlinear term in (%
\ref{a002}) is always a ``friction term'', independent of the sign of $a,$
and the case of $n$ odd will therefore turn out to be easier to treat than
the case of $n$ even. It is well known \cite{kupi} that for $n>3$ and small
bounded integrable initial conditions, the nonlinearity in (\ref{a002})
becomes irrelevant for large times in the sense that the solution converges
to a multiple of $\exp (-x^{2}/4t)/\sqrt{t},$ which solves the linear
equation $a_{t}=a_{xx}.$ We would therefore expect that, for $n>3,$ the
function $F$ is of the order $\mathcal{O}(t^{-n/2})$ on the diffusive scale.
This turns out to be wrong. As we will prove below, $F$ is of the order $%
\mathcal{O}(t^{-n/(n-1)})$ for $n>3,$ because $F$ converges on this scale
pointwise to a function that is not integrable at the origin. This
corresponds to a solution of (\ref{a002}) for which the nonlinear term is a
marginal perturbation, i.e., a solution with an amplitude of the order $%
\mathcal{O}(t^{-1/(n-1)})$. We will see that one can take advantage of this
fact, and a diffusive stability bound will be good enough to prove
convergence of $F$ to its limit, but as a consequence, our results will be
limited to the case $n>3.$ The cases $n=2$ and $n=3$ are special and will
not be discussed any further.

The following theorem is our main result.

\begin{theorem}
\label{main}For arbitrary but fixed $n\in\mathbf{N,}$ $n\geq4,$ there exist $%
\tau>0,$ functions $\mu_{1,}$ $\mu_{2},$ $\varphi_{1},$ $\varphi_{2},$ and a
class of initial conditions (specified at $t=\tau),$ such that (\ref{eqv})
has a unique solution $v$ that satisfies for all $t\geq\tau$ the bound 
\begin{equation}
\left| v(x,t)-v_{\infty}(x,t)\right| <\frac{\mathrm{const.}}{t^{4\gamma}}~,
\label{bound}
\end{equation}
where 
\begin{equation}
v_{\infty}(x,t)=\mu_{1}(\frac{\left| x\right| }{\sqrt{t}})+t^{-\varepsilon
}\mu_{2}(\frac{\left| x\right| }{\sqrt{t}})+t^{-\gamma}\varphi_{1}(\frac{%
\left| x\right| }{t^{\alpha}})+t^{-3\gamma}\varphi_{2}(\frac{\left| x\right| 
}{t^{\alpha}})~,  \label{vinfty}
\end{equation}
$\gamma=\frac{1}{2n+1},$ $\varepsilon=\frac{1}{n-1}$ and $\alpha=\frac{1}{2}%
-\gamma.$
\end{theorem}

\begin{remark}
This theorem is a local result, in the sense that the class of initial
conditions will be a set of functions in a (small) neighborhood of the
function $v_{\infty,0},$ $v_{\infty,0}(x)=v_{\infty}(x,\tau).$ In
particular, our methods do not allow us to show that the solution with the
``natural'' initial condition $v_{0}\equiv1$ at $t=0$ belongs to this set at 
$t=\tau.$ We do expect, however, that this is the case, as has been proved
for $n=1$ in \cite{Wittwer}.
\end{remark}

\begin{remark}
We note that, if an initial condition $v_{0}$ is such that $v_{0}(x)-\left|
u(x/\sqrt{\tau})\right| <0$ for a certain $x,$ then $a_{0}(x)<0$, if $x>0,$
or $b_{0}(x)<0$ if $x<0.$ A priori, we do not need to consider such initial
conditions, since in our model $a$ and $b$ represent particle densities, and
the solutions $a$ and $b$ are positive if the initial conditions $a_{0}$ and 
$b_{0}$ are positive. As we will see, for $n\geq4,$ it will not be necessary
to impose that $a_{0}$ and $b_{0}$ be positive everywhere, and it will
neither be necessary to impose that $v_{0}=a_{0}+b_{0}$ be an even function.
\end{remark}

As we will see, the functions $\varphi _{1}$ and $\varphi _{2}$ are small on
the diffusive scale, i.e., for $x\approx \sqrt{t}y,$ $y\neq 0,$ and $t$
large, 
\begin{equation}
v_{\infty }(\sqrt{t}y,t)=\mu _{1}(\left| y\right| )+t^{-\varepsilon }\mu
_{2}(\left| y\right| )+\mathcal{O}(t^{-2\varepsilon ^{\prime }})~,
\label{asym001}
\end{equation}
where $\varepsilon ^{\prime }=\varepsilon $ if $n>5,$ and $2\gamma
<\varepsilon ^{\prime }<\varepsilon $ if $n=4,$ $5.$ Using the definition (%
\ref{defv}), (\ref{defu}) for $v$ and $u,$ we therefore find that for $y>0$
and $t$ large, 
\begin{equation*}
a(\sqrt{t}y,t)=\frac{1}{2}t^{-\varepsilon }\mu _{2}(y)+\mathcal{O}%
(t^{-2\varepsilon ^{\prime }})~,
\end{equation*}
and similarly for $b,$ for $y<0.$ In contrast to the case $n=1,$ where only
exponentially few particles reach the diffusive scale, the amount of
particles decays only slowly for $n>3.$ Our results imply that, for large
times, the density of the remaining particles is given by the function $\mu
_{2},$ i.e., it is independent of the initial conditions.

As a corollary to Theorem \ref{main} we get a precise description of the
reaction front $F$ on the reactive and the diffusive scale. This description
will be given in Section 4, once we have defined the functions $\mu _{1},$ $%
\mu _{2},$ $\varphi _{1}$ and $\varphi _{2}$ in Section 3. In Section 2 we
explain our strategy for proving Theorem \ref{main}. This strategy is
implemented in Section 5 and Section 6. The Appendix contains the proof of
the existence of the functions $\mu _{2},$ $\varphi _{1}$ and $\varphi _{2}$.

\section{Strategy of the Proof}

Consider functions $v$ of the form 
\begin{equation}
v(x,t)=v_{\infty}(x,t)+\psi(x,t)~,  \label{ansatz}
\end{equation}
with $v_{\infty}$ as in Theorem \ref{main}, and $\psi(x,\tau)=\psi_{0}(x),$
for some $\tau>>1,$ with $\psi_{0}\in L_{1}\cap L_{\infty}.$ Substituting (%
\ref{ansatz}) into (\ref{eqv}) leads to an equation for the function $\psi$
of the form 
\begin{equation}
\dot{\psi}=\psi^{\prime\prime}-V\psi-I-T(\psi)~,  \label{eqpsi}
\end{equation}
for certain functions $V$ and $I,$ and for $T$ some nonlinear map. We will
show that if $v_{\infty}$ is defined correctly, $\tau$ large enough and $%
\psi_{0}$ small enough, then $V$ can be chosen positive and $T$ will be
small, so that the solution of equation (\ref{eqpsi}) will be bounded for
large times by the corresponding solution of the inhomogeneous heat equation 
$\dot{\psi}=\psi^{\prime\prime}-I.$ We will find that, with the right choice
of $v_{\infty},$%
\begin{equation}
\int dx~\left| I(\sqrt{t}x,t)\right| \leq\mathrm{const.}\text{ }%
t^{-1-4\gamma}~,  \label{boundi}
\end{equation}
from which the bound (\ref{bound}) will follow. We note that $4\gamma<\frac
{1}{2}$ for $n\geq4>\frac{7}{2},$ so that contributions of initial
conditions will become irrelevant for large times, i.e., the solution $\psi$
becomes what is called ``slaved to the inhomogeneous term''.

\section{Asymptotic Expansion}

In order to implement the strategy outlined in Section 2, we need a function 
$v_{\infty }$ that approximates the solution $v$ for large times
sufficiently well, uniformly in $x.$ Since we would like to control the time
evolution of equation (\ref{eqpsi}) on $L_{1}\cap L_{\infty },$ this
function $v_{\infty }$ needs to satisfy $\lim_{x\rightarrow \pm \infty
}v_{\infty }(x,t)=1$ in order for $v$ to satisfy the boundary conditions (%
\ref{limitv}). Furthermore, the inhomogeneous term $I$ in equation (\ref
{eqpsi}) contains second derivatives of $v_{\infty },$ and the function $I$
can therefore only be in $L_{1}\cap L_{\infty }$ if $v_{\infty }$ is at
least twice differentiable. We now construct a function $v_{\infty }$
satisfying these requirements through a two length--scale asymptotic
expansion.

To simplify the notation later on we use the convention that, unless stated
otherwise, 
\begin{equation}
y\equiv\frac{x}{\sqrt{t}}~,  \label{defy}
\end{equation}
and 
\begin{equation}
z\equiv\frac{x}{t^{\alpha}}~,  \label{defz}
\end{equation}
and we will refer to $y$ as the diffusive length scale and to $z$ as the
reactive length scale.

The function $v_{\infty}$ is given by the first and second order terms of a
so called ``matched asymptotic expansion''. The ``matched'' refers to the
fact that such an expansion contains functions that can not be associated
uniquely with one of the length scales and can therefore be used to
``match'' the behavior at large distances of the shorter scale with the
behavior at short distances of the larger scale. Let 
\begin{equation}
\delta=\frac{n+2}{n-1}~,  \label{defdelta007}
\end{equation}
and let $\gamma,$ $\alpha,$ and $\varepsilon$ be as in Theorem \ref{main}.
Then, the functions $y\equiv t^{-\gamma}z,$ $t^{-\varepsilon}y^{-\delta}%
\equiv t^{-\gamma}z^{-\delta},$ $t^{-\varepsilon}y^{2-\delta}\equiv
t^{-3\gamma }z^{2-\delta}$ and $t^{-2\varepsilon}y^{-2\delta+1}\equiv
t^{-3\gamma }z^{-2\delta+1}$ are of this form and will naturally show up in
the function $v_{\infty}.$ As a consequence, the representation (\ref{vinfty}%
) for $v_{\infty}$ is not unique. If we choose (as we will) to compute the
expansion for $v_{\infty}$ in the order of decreasing amplitudes, i.e., if
we first compute the term of order $\mathcal{O}(t^{-\gamma}),$ then the term
of order $\mathcal{O}(t^{-\varepsilon}),$ and finally the term of order $%
\mathcal{O}(t^{-3\gamma}),$ we get a representation of $v_{\infty}$ of the
form 
\begin{equation}
v_{\infty}(x,t)=\mu_{1}(\left| y\right| )+t^{-\gamma}\eta(\left| z\right|
)+t^{-\varepsilon}\left( \mu_{2}(\left| y\right| )-\lambda\left| y\right|
^{-\delta}\right) +t^{-3\gamma}\varphi_{2}(\left| z\right| )~,
\label{repres1}
\end{equation}
where 
\begin{equation}
\eta(z)=\varphi_{1}(z)+\lambda z^{-\delta}~,  \label{etaphi1}
\end{equation}
with $\lambda$ a certain constant to be determined later.

We note that, by definition, $v_{\infty}$ is a symmetric function, and it is
therefore sufficient to consider positive values of $x$ if we choose
appropriate boundary conditions at $x=0$ to ensure regularity.

Finally, since we will need to describe the asymptotic behavior of various
functions near zero and infinity, we introduce the following notation. Let $%
f $ be a continuous function from $\mathbf{R}_{+}$ to $\mathbf{R,}$ $k$ a
positive integer and $p_{1}<p_{2}<\dots<p_{k}$ real numbers. Then, we say
that 
\begin{equation*}
f(x)=\sum_{i=1}^{k}f_{i}~x^{p_{i}}+\dots
\end{equation*}
near $x=0,$ if 
\begin{equation*}
\lim_{x\rightarrow0^{+}}\frac{1}{x^{p_{k}}}\left|
f(x)-\sum_{i=1}^{k}f_{i}~x^{p_{i}}\right| =0~,
\end{equation*}
and we say that 
\begin{equation*}
f(x)=\sum_{i=1}^{k}f_{i}~x^{-p_{i}}+\dots
\end{equation*}
near $x=\infty,$ if 
\begin{equation*}
\lim_{x\rightarrow\infty}x^{p_{k}}\left|
f(x)-\sum_{i=1}^{k}f_{i}~x^{-p_{i}}\right| =0~.
\end{equation*}

\subsection{Equation for $\protect\mu_{1}$}

To lowest order the function $v$ is asymptotic to $\mu_{1}(\left| y\right| ) 
$, with $\mu_{1}$ as defined in (\ref{defmu}). We note that $\mu_{1}$ has
near $y=0$ the expansion 
\begin{equation}
\mu_{1}(y)=\kappa y+\kappa_{3}y^{3}+\dots~,  \label{mu1nearzero}
\end{equation}
where $\kappa=\frac{1}{\sqrt{\pi}}$ and $\kappa_{3}=-\frac{1}{12}\kappa$.
Furthermore, $\lim_{y\rightarrow\infty}\mu_{1}(y)=1.$

\subsection{Equation for $\protect\varphi_{1}$}

We make the ansatz $v(x,t)=\mu_{1}(\left| y\right| )+t^{-\gamma}\eta(\left|
z\right| )$ which we substitute into equation (\ref{eqv}). We multiply the
resulting equation with $t^{\gamma+2\alpha},$ and take then the limit $%
t\rightarrow\infty,$ keeping $z$ fixed. This leads to the differential
equation 
\begin{equation}
\eta^{\prime\prime}=\left( 2\kappa z\eta+\eta^{2}\right) ^{n}~,
\label{eqeta}
\end{equation}
where $z$ is now considered a variable in $\mathbf{R}_{+}.$ Since $%
\lim_{y\rightarrow\infty}\mu_{1}(y)=1,$ the correct boundary condition for $%
\eta$ at infinity is 
\begin{equation}
\lim_{z\rightarrow\infty}\eta(z)=0~,  \label{etainfinity}
\end{equation}
and at $z=0$ we impose 
\begin{equation}
\eta^{\prime}(0)=-\kappa~,  \label{etazero}
\end{equation}
which makes the function $\mu_{1}(\left| y\right| )+t^{-\gamma}\eta(\left|
z\right| )$ twice differentiable at $x=0,$ since $\mu_{1}$ and $\eta$ are
twice differentiable at zero and $\partial_{x}\left( \mu_{1}(\left| y\right|
)+t^{-\gamma}\eta(\left| z\right| )\right) (0)=0.$ A proof of the following
proposition can be found in the appendix.

\begin{proposition}
\label{theoremeta}For $n\in \mathbf{N,}$ $n\geq 2,$ there exists a unique
function $\eta \colon \mathbf{R}_{+}\!\rightarrow \mathbf{R}$ that satisfies
equation (\ref{eqeta}) with the boundary conditions (\ref{etainfinity}) and (%
\ref{etazero}). The function $\eta $ is positive, and has near $z=0$ the
expansion 
\begin{equation*}
\eta (z)=\eta _{0}-\kappa z+\eta _{2}z^{2}-\eta _{4}z^{4}+\dots ~,
\end{equation*}
with positive coefficients $\eta _{0},$ $\eta _{2}$ and $\eta _{4}.$ For $z$
large, $\eta $ is of the form 
\begin{equation*}
\eta (z)=\frac{\lambda }{z^{\delta }}+\frac{\lambda _{\infty }}{z^{\delta
^{\prime }}}+\dots ~,
\end{equation*}
for a certain constant $\lambda _{\infty },$ with $\delta $ as in (\ref
{defdelta007}), 
\begin{equation}
\lambda =\left( \frac{\delta (\delta +1)}{(2\kappa )^{n}}\right) ^{1/(n-1)}~,
\label{deflambda007}
\end{equation}
and 
\begin{equation*}
\delta ^{\prime }=\left\{ 
\begin{array}{ccc}
\left( \sqrt{4n\delta (\delta +1)+1}-1\right) /2 &  & 2\leq n\leq 5~, \\ 
&  &  \\ 
2\delta +1 &  & n\geq 6~.
\end{array}
\right.
\end{equation*}
\end{proposition}

We note that $3<\delta^{\prime}\leq2\delta+1.$ The constants $\eta_{0},$ $%
\eta_{2},$ $\eta_{4}$ and $\lambda_{\infty}$ are given in the appendix. The
function $\varphi_{1}$ is defined in (\ref{etaphi1}) in terms of $\eta.$

\subsection{Equation for $\protect\mu_{2}\label{smu2}$}

We make the ansatz $v(x,t)=\mu_{1}(\left| y\right| )+t^{-\gamma}\eta(\left|
z\right| )+t^{-\varepsilon}(\mu_{2}(\left| y\right| )-\lambda\left| y\right|
^{-\delta})$ which we substitute into equation (\ref{eqv}). We multiply the
resulting equation with $t^{1+\varepsilon},$ and take then the limit $%
t\rightarrow\infty,$ keeping $y$ fixed. Since $\eta(z)=\eta(t^{\gamma }y)$
and $\lim_{t\rightarrow\infty}t^{\varepsilon-\gamma}\eta(z)-\lambda
y^{-\delta}=0,$ this leads to the differential equation for the function $%
\mu_{2},$ 
\begin{equation}
\mu_{2}^{\prime\prime}+\frac{1}{2}y\mu_{2}^{\prime}+\varepsilon\mu_{2}=(2%
\mu_{1}\mu_{2})^{n}~,  \label{mu2}
\end{equation}
where $y$ is now considered as a variable in $\mathbf{R}_{+}.$ At $y=0$ we
impose the boundary condition 
\begin{equation}
\lim_{y\rightarrow0}\mu_{2}(y)y^{\delta}=\lambda~,  \label{bmu2-1}
\end{equation}
which removes the leading singularity of the function $\mu_{2}(\left|
y\right| )-\lambda\left| y\right| ^{-\delta}$ at $y=0.$ As we will see, the
sub--leading singularity is proportional to $\left| y\right| ^{2-\delta},$
which is not a twice differentiable function at $y=0$ (except for $n=4$
where $\delta=2).$ This singularity will be cancelled by imposing
appropriate boundary conditions for the function $\varphi_{2}.$

The correct choice of boundary conditions for $\mu _{2}$ at infinity is
somewhat less obvious. In the appendix we show that the condition $%
\lim_{y\rightarrow \infty }\mu _{2}(y)=0$ is not sufficient to single out a
unique function $\mu _{2}.$ If $\mu _{2}$ does converge to zero at infinity,
then it is asymptotic to a solution of the equation 
\begin{equation*}
\mu ^{\prime \prime }+\frac{1}{2}y\mu ^{\prime }+\varepsilon \mu =0~.
\end{equation*}
This linear equation is compatible with a (very slow) algebraic decay, $\mu
_{2}(y)\approx y^{-2\varepsilon },$ or with a modified Gaussian decay, $\mu
_{2}(y)\approx \exp (-y^{2}/4)/y^{1-2\varepsilon },$ with the algebraic
decay being the generic case. It will be essential in later sections that $%
\mu _{2}$ decays rapidly at infinity, and we therefore impose the boundary
condition 
\begin{equation}
\lim_{y\rightarrow \infty }\mu _{2}(y)y^{2\varepsilon }=0~.  \label{bmu2-2}
\end{equation}
A proof of the following proposition can be found in the appendix.

\begin{proposition}
\label{cmu2}For all $n\geq 4,$ there exists a unique positive function $\mu
_{2}\colon \mathbf{R}_{+}\rightarrow \mathbf{R}$ that satisfies equation (%
\ref{mu2}) with the boundary conditions (\ref{bmu2-1}) and (\ref{bmu2-2}).
For $y$ small, the function $\mu _{2}$ is of the form 
\begin{equation}
\mu _{2}(y)=\lambda y^{-\delta }+\lambda _{0}y^{2-\delta }+\lambda
_{1}y^{4-\delta }+\dots ~,  \label{asymmu1}
\end{equation}
with 
\begin{equation}
\lambda _{0}=\frac{1}{2}\frac{\lambda }{\kappa }\frac{-2n\kappa _{3}\delta
(\delta +1)-\kappa (\delta -2\varepsilon )}{\left( n-1\right) \delta (\delta
+1)+2(2\delta -1)}>0~,  \label{lambda0}
\end{equation}
with $\lambda _{1}\neq 0$ and with $\lambda $ as in Proposition \ref
{theoremeta}. For $y$ large, the function $\mu _{2}$ decays rapidly in the
sense that 
\begin{equation}
\mu _{2}(y)=\exp (-\frac{y^{2}}{4})\left( \frac{C}{y^{1-2\varepsilon }}%
+\dots \right) ~,  \label{asymmu2}
\end{equation}
for some constant $C>0.$
\end{proposition}

\subsection{Equation for $\protect\varphi_{2}$}

We make the ansatz $v(x,t)=\mu_{1}(\left| y\right| )+t^{-\gamma}\eta(\left|
z\right| )+t^{-\varepsilon}(\mu_{2}(\left| y\right| )-\lambda\left| y\right|
^{-\delta})+t^{-3\gamma}\varphi_{2}(\left| z\right| )$ which we substitute
into equation (\ref{eqv}). We multiply the resulting equation with $%
t^{3\gamma+2\alpha},$ and take then the limit $t\rightarrow\infty,$ keeping $%
z$ fixed. This leads to the (linear) differential equation for $\varphi
_{2}, $ 
\begin{equation}
\varphi_{2}^{\prime\prime}+\gamma\eta+\alpha z\eta^{\prime}+(2-\delta
)(1-\delta)\lambda_{0}z^{-\delta}=n\left( 2\kappa z\eta+\eta^{2}\right)
^{n-1}\left[ (2\kappa
z+2\eta)(\varphi_{2}+\lambda_{0}z^{2-\delta})+2\kappa_{3}z^{3}\eta\right] ~.
\label{varphi2}
\end{equation}
In order to compensate the sub--leading singular behavior of $\mu_{2}$ near $%
x=0$ we make the ansatz 
\begin{equation}
\varphi_{2}(z)=-\lambda_{0}z^{2-\delta}+h(z)~,  \label{newh}
\end{equation}
which we substitute into equation (\ref{varphi2}). For the function $h$ we
get the equation 
\begin{equation}
h^{\prime\prime}+\gamma\eta+\alpha z\eta^{\prime}=n\left( 2\kappa z\eta
+\eta^{2}\right) ^{n-1}\left( (2\kappa z+2\eta)h+2\kappa_{3}z^{3}\eta\right)
~.  \label{h}
\end{equation}
Since the function $\eta$ is regular near $z=0,$ the solution $h$ turns out
to be regular near $z=0,$ too. Therefore, the function $z\mapsto h(\left|
z\right| )$ is twice differentiable near $x=0$ if we impose at $z=0$ the
boundary condition 
\begin{equation}
h^{\prime}(0)=0~.  \label{bh-1}
\end{equation}
At infinity we need that $\lim_{z\rightarrow\infty}\varphi_{2}(z)=0.$ We
therefore require that 
\begin{equation}
\lim_{z\rightarrow\infty}(h(z)-\lambda_{0}z^{2-\delta})=0~.  \label{bh-2}
\end{equation}
A proof of the following proposition can be found in the appendix.

\begin{proposition}
\label{cfi22}For all $n\geq 4,$ there exists a unique function $h\colon 
\mathbf{R}_{+}\rightarrow \mathbf{R}$ that satisfies equation (\ref{h}) with
the boundary conditions (\ref{bh-1}) and (\ref{bh-2}). Near $z=0,$ the
function $h$ is of the form 
\begin{equation*}
h(z)=h_{0}+h_{2}z^{2}+\dots ~,
\end{equation*}
with certain coefficients $h_{0}$ and $h_{2},$ and for $z$ large $h$ is of
the form 
\begin{equation*}
h(z)=\lambda _{0}z^{2-\delta }+\frac{\lambda ^{\prime }}{z^{\delta ^{\prime
}-2}}+\dots ~,
\end{equation*}
with $\lambda _{0}$ as defined in (\ref{lambda0}), for a certain constant $%
\lambda ^{\prime }$, and with $\delta ^{\prime }$ as defined in Proposition 
\ref{theoremeta}.
\end{proposition}

\section{The Reaction Front}

Using the properties of the functions $\mu_{1},$ $\mu_{2},$ $\varphi_{1}$
and $\varphi_{2},$ we get from Theorem \ref{main} the following behavior of
the reaction front $F.$

\begin{corollary}
Let $v$ be as in Theorem \ref{main}, and $F$ as defined in (\ref{reaction001}%
). Then, for all $z\in \mathbf{R,}$%
\begin{equation*}
\lim_{t\rightarrow \infty }t^{2n\gamma }F(t^{\alpha }z,t)=\frac{1}{2}%
(2\kappa \left| z\right| \eta (\left| z\right| )+\eta (\left| z\right|
)^{2})^{n}=\frac{1}{2}\eta ^{\prime \prime }(\left| z\right| )=\left\{ 
\begin{array}{lll}
\eta _{2}-6\eta _{4}\left| z\right| ^{2}+\dots & \text{for} & \noindent
\left| z\right| \approx 0~, \\ 
&  &  \\ 
\frac{1}{2}\left( 2\lambda \kappa \right) ^{n}/\left| z\right| ^{\delta
+2}+\dots & \text{for} & \noindent \left| z\right| >>1~,
\end{array}
\right.
\end{equation*}
and for all $y\neq 0,$%
\begin{equation*}
\lim_{t\rightarrow \infty }t^{n\varepsilon }F(\sqrt{t}y,t)=\frac{1}{2}(2\mu
_{1}~\mu _{2})^{n}(\left| y\right| )=\left\{ 
\begin{array}{lll}
\frac{1}{2}\left( 2\lambda \kappa \right) ^{n}/\left| y\right| ^{\delta
+2}+\dots & \text{for} & \left| y\right| \approx 0~, \\ 
&  &  \\ 
\exp (-n\left| y\right| ^{2}/4)(2^{n-1}C^{n}/\left| y\right|
^{n(1-2\varepsilon )}+\dots ) & \text{for} & \left| y\right| >>1~.
\end{array}
\right.
\end{equation*}
Here, $\eta _{2},$ $\eta _{4}$ are as defined in Proposition \ref{theoremeta}
and $C$ is as defined in (\ref{asymmu2}).
\end{corollary}

\section{The Equation for $\protect\psi$}

In order to simplify the notation we define the function $\overline{u},$ 
\begin{equation*}
\overline{u}(x,t)=\mu_{1}(\left| y\right| )~,
\end{equation*}
the function $\mu_{3},$%
\begin{equation*}
\mu_{3}(y)=\mu_{2}(y)-\lambda y^{-\delta}~,
\end{equation*}
the function $\phi,$%
\begin{equation}
\phi(x,t)=t^{-\gamma}\eta(|z|)+t^{-\varepsilon}\mu_{3}(|y|)+t^{-3\gamma
}\varphi_{2}(|z|)~,  \label{defphi}
\end{equation}
and the function $\phi_{1},$%
\begin{equation}
\phi_{1}(x,t)=\phi(x,t)-\kappa\frac{\left| x\right| }{\sqrt{t}}~.
\label{defphi1}
\end{equation}
The function $v_{\infty}$ in Theorem \ref{main} and in (\ref{repres1}) can
then be written as $v_{\infty}=\overline{u}+\phi.$

Let now $v=v_{\infty}+\psi.$ Then, 
\begin{align*}
\left( v^{2}-u^{2}\right) ^{n} & =\left( \left( \overline{u}+\phi
+\psi\right) ^{2}-u^{2}\right) ^{n}=\left( \left( 2\overline{u}\phi
+\phi^{2}\right) +\left( 2(\overline{u}+\phi)+\psi\right) \psi\right) ^{n} \\
& =\sum_{k=0}^{n}\binom{n}{k}\left( 2\overline{u}\phi+\phi^{2}\right)
^{n-k}\left( 2(\overline{u}+\phi)+\psi\right) ^{k}\psi^{k}~.
\end{align*}
Therefore, substituting the ansatz $v=v_{\infty}+\psi$ into (\ref{eqv})
leads to the following equation for the function $\psi,$%
\begin{equation}
\dot{\psi}=\psi^{\prime\prime}-\widehat{V}\psi-I-\widehat{T}(\psi )~,
\label{equpsi1}
\end{equation}
with the function $\widehat{V},$ 
\begin{equation}
\widehat{V}=2n~(2\overline{u}\phi+\phi^{2})^{n-1}~(\overline{u}+\phi )~,
\label{vhat}
\end{equation}
the function $I,$%
\begin{equation}
I=-\dot{\phi}+\phi_{1}^{\prime\prime}-\left( 2\overline{u}\phi+\phi
^{2}\right) ^{n}~,  \label{equi}
\end{equation}
and the map $\widehat{T},$%
\begin{equation}
\widehat{T}(\psi)=n\left( 2\overline{u}\phi+\phi^{2}\right) ^{n-1}\psi
^{2}+\sum_{k=2}^{n}\binom{n}{k}\left( 2\overline{u}\phi+\phi^{2}\right)
^{n-k}\left( 2(\overline{u}+\phi)+\psi\right) ^{k}\psi^{k}~.  \label{that}
\end{equation}

\subsection{The function $V\label{sectionv}$}

The function $\widetilde{\mu},$ $\widetilde{\mu}(y)=\mu_{1}(y)/y$ is
strictly decreasing on $\mathbf{R}_{+},$ and therefore $\mu_{1}(y)/y\geq\mu
_{1}(t^{\gamma}y)/(t^{\gamma}y)$ for $t\geq1.$ Furthermore, the functions $%
\eta$ and $\mu_{1}$ are strictly positive and $\mu_{1}$ is strictly
increasing. These properties imply that, for $t\geq\tau\geq1,$ $\overline
{u}(x,t)+t^{-\gamma}\eta(\left| z\right| )=$ $\mu_{1}(\left| y\right|
)+t^{-\gamma}\eta(\left| z\right| )\geq$ $t^{-\gamma}(\mu_{1}(\left|
z\right| )+\eta(\left| z\right| ))$ $\geq t^{-\gamma}c_{0}>0,$ where $%
c_{0}=\inf_{z>0}(\mu_{1}(z)+\eta(z)).$ Next, since the functions $\mu_{3}$
and $\varphi_{2}$ are bounded and since $3\gamma\geq\varepsilon,$ for $%
n\geq4,$ we have that $\left| t^{-\varepsilon}\mu_{3}(\left| y\right|
)+t^{-3\gamma}\varphi_{2}(\left| z\right| )\right| <\mathrm{const.~}%
t^{-\varepsilon},$ and as a consequence $(\overline{u}+\phi)$ and $(2%
\overline{u}+\phi)$ are positive functions of $x$ for all fixed $t\geq
\tau_{0},$ if $\tau_{0}$ large enough.

\begin{proposition}
\label{propvodd}For $n$ odd, $n\geq 5,$ there exists $\tau _{0}\geq 1,$ such
that for all $t\geq \tau _{0}$ the function $\widehat{V}$ is positive.
\end{proposition}

\begin{proof}
The function $(2\overline{u}\phi +\phi ^{2})^{n-1}$ is positive, for $n$ odd.
\end{proof}

\medskip

As a consequence, for $n$ odd, equation (\ref{equpsi1}) is of the form
indicated in Section 2, with $V=\widehat{V}$ and $T=\widehat{T}.$ The rest
of this section treats the case of $n$ even, which, as indicated in the
introduction, is slightly more delicate. It can be skipped in a first
reading or if the reader is only interested in the case of $n$ odd.

So let $n$ be even. The idea is to split $\widehat{V}$ into its positive
part $V=$ $\widehat{V}_{+}$ and its negative part $V_{1}=\widehat{V}_{-}~$,
and to show that $\widehat{V}_{-}$ is small enough so that it can be treated
together with the nonlinear term. Consider the function $\phi $ defined in (%
\ref{defphi}). The problem is that $\phi $ becomes negative for large values
of $x$, and that therefore $\widehat{V}$ becomes negative for large values
of $x.$ To understand why $\phi $ becomes negative, we note that the leading
order term $t^{-\gamma }\lambda z^{-\delta }$ in the large $z$ asymptotics
of $t^{-\gamma }\eta (z)$ is compensated by the leading order term $%
-t^{-\varepsilon }\lambda y^{-\delta }$ in the large $y$ asymptotics of $%
t^{-\varepsilon }\mu _{3}(y).$ The leading order of $\phi $ at $x$ large is
therefore given by the second order term in the large $z$ asymptotics of $%
\eta $ and the leading term in the large $z$ asymptotics of $\varphi _{2}.$
The first of these terms is proportional to $t^{-\gamma }z^{-\delta ^{\prime
}},$ and the second one is proportional to $t^{-3\gamma }z^{2-\delta
^{\prime }}\equiv t^{-\gamma }y^{2}z^{-\delta ^{\prime }}.$ The
corresponding proportionality constants $\lambda _{\infty }$ and $\lambda
^{\prime }$ can be computed for $n>5$ and turn out to be negative. For $%
n=4,5 $ these constants can not be obtained from asymptotic expansions, but
numerical results show that they are in fact also negative in these cases.
We do not need a proof of this numerical fact, because the following
proposition is also correct for positive $\widehat{V}$.

\begin{proposition}
\label{propV}For $n$ even, $n\geq4,$ there exists $\tau_{1}\geq1$, such that
the function $V_{1}$, satisfies for all $t\geq\tau_{1}$ the bound 
\begin{equation}
\sup_{x\in\mathbf{R}}|V_{1}(x,t)|\leq\mathrm{const.}~t^{-\gamma(n-1)(\delta
^{\prime}+1)}~.  \label{v1bound}
\end{equation}
\end{proposition}

\begin{proof}
The idea is to write $\phi $ as the sum of a function $\phi _{0}$ that is
positive and a function $\phi _{\infty }$ that absorbs the asymptotic
behavior at infinity. Since $\mu _{3}(y)\approx \lambda _{0}y^{2-\delta }$
for $y$ small, with $\lambda _{0}>0,$ there exists $y_{0}>0$ such that $\mu
_{3}(|y|)\geq 0,$ for all $|y|\leq y_{0}.$ Let $c>0,$ to be chosen below,
and let $\theta $ be the Heaviside step function, i.e., $\theta (x)=1$ for $%
x>0,$ and $\theta (x)=0$ for $x<0.$ Then, we define the function $\phi
_{\infty }$ by the equation 
\begin{equation*}
\phi _{\infty }(x,t)=-ct^{-\gamma }\theta (\left| y\right| -y_{0})\left|
y\right| ^{2}\left| z\right| ^{-\delta ^{\prime }}~,
\end{equation*}
and we set $\phi _{0}=\phi -\phi _{\infty }.$ In order to prove that $\phi
_{0}$ is positive, for $c$ large enough, we write $\phi _{0}=\phi
_{0}^{\left( 1\right) }+\phi _{0}^{\left( 2\right) },$ where 
\begin{equation*}
\phi _{0}^{(1)}(x,t)=t^{-\gamma }(\eta (\left| z\right| )-\lambda
|z|^{-\delta }\theta (\left| y\right| -y_{0}))+t^{-3\gamma }\varphi
_{2}(\left| z\right| )+c~\theta (\left| y\right| -y_{0})t^{-3\gamma
}|z|^{2-\delta ^{\prime }}~,
\end{equation*}
and 
\begin{equation*}
\phi _{0}^{(2)}(x,t)=t^{-\varepsilon }\left( \mu _{3}(\left| y\right|
)+\lambda |y|^{-\delta }\theta (\left| y\right| -y_{0})\right) ~.
\end{equation*}
$\phi _{0}^{(2)}$ is positive for $\left| y\right| >y_{0},$ since in this
case $\phi _{0}^{(2)}(x,t)=\mu _{2}(\left| y\right| )>0,$ and $\phi
_{0}^{(2)}$ is positive for $\left| y\right| <y_{0}$ by definition of $%
y_{0}. $ Next we consider $\phi _{0}^{(1)}.$ For $\left| z\right| <t^{\gamma
}y_{0}$ we have that $\phi _{0}^{(1)}(x,t)=t^{-\gamma }\eta (\left| z\right|
)+t^{-3\gamma }\varphi _{2}(\left| z\right| )$. But $t^{-\gamma }\eta
(z)+t^{-3\gamma }\varphi _{2}(z)>0$ for all $z\geq 0,$ and all $t\geq \tau ,$
if $\tau $ is sufficiently large, since $\eta >0,$ since $\varphi _{2}$ is
bounded, and since $\left| \varphi _{2}(z)\right| <\eta (z)$ for $z$ large
enough. Finally, using the asymptotic properties of $\eta $ and $\varphi
_{2} $ we see that $\phi _{0}^{(1)}>0$ for $\left| z\right| >t^{\gamma
}y_{0} $ if $c$ is chosen large enough.

We now estimate the function $V_{1}.$ From the definition of $\phi _{\infty
} $ we get that 
\begin{equation*}
|\phi _{\infty }(x,t)|\leq \mathrm{const.}\text{ }t^{-\gamma (\delta
^{\prime }+1)}~,
\end{equation*}
and therefore, since $\phi _{0}$ is positive, we have the lower bound 
\begin{equation*}
\phi (x,t)^{n-1}\geq \overline{c}~t^{-\gamma (n-1)(\delta ^{\prime }+1)}~,
\end{equation*}
for some constant $\overline{c}<0$, from which (\ref{v1bound}) follows.
\end{proof}

\subsection{The function $I$}

\begin{theorem}
\label{thm:Inhomo} Let $I$ be as defined in (\ref{equi}), and let $n\geq4.$
Then, there exists a constant $c_{I}>0,$ such that for all $t\geq1,$%
\begin{equation}
\int_{-\infty}^{\infty}dx~|I(\sqrt{t}x,t)|\leq c_{I}~t^{-1-4\gamma }~.
\label{eqn:thminhomo}
\end{equation}
\end{theorem}

The function $I$ is even, and it is therefore sufficient to bound it for $%
x\geq0.$ The strategy of the proof will be to rewrite the function $I$ as a
sum of functions of the form $t^{-\sigma}G(y)F(z),$ with $\sigma>0,$ and
with $G$ and $F$ functions with appropriate asymptotic behavior at zero and
infinity. Each of the terms in the sum can then be estimated with the help
of Lemma \ref{lem:estimations} below. In order to keep the notation as
simple as possible, we suppress in what follows the arguments of functions
whenever there is no risk of confusion.

\begin{proposition}
For $x\geq0,$ the function $-I$ is of the form 
\begin{equation}
-I=\sum_{p=2}^{n}\sum_{q=0}^{p}A_{p,q}+\sum_{i=2}^{8}A_{i}~,  \label{repi}
\end{equation}
where 
\begin{align*}
A_{2} & =\sum_{p=1}^{n-1}\binom{n-1}{p}t^{-2\gamma(n-1-p)-p%
\varepsilon}T_{1}^{n-1-p}T_{2}^{p}\left( t^{-4\gamma}T_{3}\right) ~, \\
A_{3} & =n(t^{-2\gamma}T_{1}+t^{-\varepsilon}T_{2})^{n-1}\left(
t^{-\gamma(1+\delta^{\prime})}T_{4}+t^{-2\varepsilon}T_{5}+t^{-6\gamma}T_{6}%
\right) ~, \\
A_{4} & =\sum_{p=2}^{n}\binom{n}{p}(t^{-2\gamma}T_{1}+t^{-%
\varepsilon}T_{2})^{n-p}(t^{-4\gamma}T_{3}+t^{-\gamma(1+\delta^{%
\prime})}T_{4}+t^{-2\varepsilon}T_{5}+t^{-6\gamma}T_{6})^{p}~, \\
A_{5} & =-t^{-1-3\gamma}\left( 3\gamma\varphi_{2}+\alpha z\varphi
_{2}^{\prime}\right) ~, \\
A_{6} & =-t^{-2n\gamma-2\gamma}nT_{1}^{n-1}2z^{2-\delta}\eta\left(
\lambda_{0}-\mu_{3}y^{\delta-2}\right) ~, \\
A_{7} & =t^{-2n\gamma+2\gamma-\varepsilon}nT_{1}^{n-1}(2\mu_{1}\mu
_{3}-2\kappa\lambda_{0}y^{3-\delta})-t^{-2n\gamma+2\gamma-\varepsilon
}n(2\kappa\lambda z^{1-\delta})^{n-1}(2\mu_{1}\mu_{3}-2\kappa\lambda
_{0}y^{3-\delta})~, \\
A_{8} & =t^{-2n\gamma+\gamma}nT_{1}^{n-1}2\eta((\mu_{1}-\kappa y)-\kappa
_{3}y^{3})-t^{-2n\gamma+\gamma}n(2\kappa\lambda z^{1-\delta})^{n-1}2\lambda
z^{-\delta}(\mu_{1}-\kappa y-\kappa_{3}y^{3})~, \\
A_{p,q} & =\binom{n}{p}\binom{p}{q}\left( R_{2}^{p,q}-R_{1}^{p,q}\right) ~,
\end{align*}
where 
\begin{align*}
R_{1}^{p,q} & =t^{-2n\gamma+2p\gamma-p\varepsilon}(2\kappa\lambda
z^{1-\delta})^{n-p}(2(\mu_{1}-\kappa y)\lambda
y^{-\delta})^{p-q}(2\mu_{1}\mu_{3})^{q}~, \\
R_{2}^{p,q} & =t^{-2n\gamma+2p\gamma-p\varepsilon}T_{1}^{n-p}(2(\mu
_{1}-\kappa y)y^{-\delta}z^{\delta}\eta)^{p-q}(2\mu_{1}\mu_{3})^{q}~,
\end{align*}
and where 
\begin{align}
T_{1}(z) & =2\kappa z\eta(z)+\eta(z)^{2}~,  \label{t1} \\
T_{2}(y,z) & =2(\mu_{1}(y)-\kappa y)y^{-\delta}z^{\delta}\eta(z)+2\mu
_{1}(y)\mu_{3}(y)~,  \label{t2} \\
T_{3}(y,z) & =\left( 2\kappa z+2\eta(z)\right) \varphi_{2}(z)+2\mu
_{3}(y)y^{\delta-2}z^{2-\delta}\eta(z)~,  \label{t3} \\
T_{4}(y,z) & =2(\mu_{1}(y)-\kappa y)y^{2-\delta^{\prime}}z^{\delta^{\prime
}-2}\varphi_{2}(z)~,  \label{t4} \\
T_{5}(y) & =\mu_{3}(y)^{2}~,  \label{t5} \\
T_{6}(y,z) & =2\mu_{3}(y)y^{\delta-2}z^{2-\delta}\varphi_{2}(z)+\varphi
_{2}(z)^{2}~.  \label{t6}
\end{align}
\end{proposition}

\begin{proof}
In terms of the functions (\ref{t1})--(\ref{t6}) we get that, for $x>0,$ 
\begin{equation*}
2\overline{u}\phi +\phi ^{2}=t^{-2\gamma }T_{1}+t^{-\varepsilon
}T_{2}+t^{-4\gamma }T_{3}+t^{-\gamma (1+\delta ^{\prime
})}T_{4}+t^{-2\varepsilon }T_{5}+t^{-6\gamma }T_{6}~,
\end{equation*}
and therefore $(2\overline{u}\phi +\phi ^{2})^{n}=B_{1}+\sum_{i=2}^{4}A_{i}$%
, where 
\begin{equation*}
B_{1}=(t^{-2\gamma }T_{1}+t^{-\varepsilon }T_{2})^{n}+n(t^{-2\gamma
}T_{1})^{n-1}(t^{-4\gamma }T_{3})~,
\end{equation*}
and where $A_{2},$ $A_{3}$ and $A_{4}$ are as defined above. Since $-I=\dot{%
\phi}-\phi _{1}^{\prime \prime }+\left( 2\overline{u}\phi +\phi ^{2}\right)
^{n},$ it remains to be shown that 
\begin{equation*}
B_{1}-\phi _{1}^{\prime \prime }+\dot{\phi}=\sum_{p=2}^{n}%
\sum_{q=0}^{p}A_{p,q}+\sum_{i=5}^{8}A_{i}~.
\end{equation*}
Using the differential equations for $\mu _{2},$ $\eta $ and $\varphi _{2},$
we find that 
\begin{equation*}
B_{1}-\phi _{1}^{\prime \prime }+\dot{\phi}=R_{1}+R_{2}+S_{3}+A_{5}~,
\end{equation*}
where $A_{5}$ as defined above, where 
\begin{align*}
R_{1}& =-t^{-n\varepsilon }\left( -(2\kappa \lambda y^{1-\delta })^{n}+(2\mu
_{1}(\lambda y^{-\delta }+\mu _{3}))^{n}\right) ~, \\
R_{2}& =(t^{-2\gamma }T_{1}+t^{-\varepsilon }T_{2})^{n}-(t^{-2\gamma
}T_{1})^{n}~,
\end{align*}
and where 
\begin{align*}
S_{3}& =-t^{-2n\gamma -2\gamma }\left[ -n(2\kappa \lambda )^{n}(\frac{\kappa
_{3}}{\kappa }+\frac{\lambda _{0}}{\lambda })z^{-\delta
}+n(T_{1})^{n-1}\left( (\varphi _{2}+\lambda _{0}z^{2-\delta })(2\kappa
z+2\eta )+2\kappa _{3}z^{3}\eta \right) \right] \\
& \hspace{3cm}+n(t^{-2\gamma }T_{1})^{n-1}(t^{-4\gamma }T_{3})~.
\end{align*}
The functions $R_{1}$ and $R_{2}$ can be further decomposed as follows 
\begin{align*}
R_{1}& =S_{1}-\sum_{p=2}^{n}\sum_{q=0}^{p}\binom{n}{p}\binom{p}{q}%
R_{1}^{p,q}~, \\
R_{2}& =S_{2}+\sum_{p=2}^{n}\sum_{q=0}^{p}\binom{n}{p}\binom{p}{q}%
R_{2}^{p,q}~,
\end{align*}
where $R_{1}^{p,q}$ and $R_{2}^{p,q}$ are as defined above, and where 
\begin{align*}
S_{1}& =-nt^{-2\gamma (n-1)-\varepsilon }(2\kappa \lambda z^{1-\delta
})^{n-1}(2(\mu _{1}-\kappa y)\lambda y^{-\delta }+2\mu _{1}\mu _{3})~, \\
S_{2}& =nt^{-2\gamma (n-1)-\varepsilon }T_{1}^{n-1}T_{2}~.
\end{align*}
It remains to be shown that 
\begin{equation*}
\sum_{i=1}^{3}S_{i}=\sum_{i=6}^{8}A_{i}~,
\end{equation*}
but this follows using the definitions.
\end{proof}

\subsubsection{Proof of Theorem \ref{thm:Inhomo}}

In order to characterize the behavior of a function near zero and infinity
we introduce the following family of vector spaces.

\begin{definition}
Let $p$ and $q$ be two real numbers with $p+q\geq0.$ Then, we define $%
\mathcal{V}(p,q)$ to be the vector space of continuous functions $F$ from $%
\mathbf{R}_{+}$ to $\mathbf{R,}$ for which the norm 
\begin{equation*}
\left\| F\right\| _{p,q}=\sup_{x\geq0}\left| F(x)\right| \left(
x^{-p}+x^{q}\right)
\end{equation*}
is finite.
\end{definition}

Note that, if a function is in $\mathcal{V}(p,q)$, then it is also in $%
\mathcal{V}(p^{\prime},q^{\prime})$ for any pair of numbers $p^{\prime}\leq
p,q^{\prime}\leq q$ for which$\ p^{\prime}+q^{\prime}\geq0.$ Furthermore, if 
$F_{1}$ is in $\mathcal{V}(p_{1},q_{1})$, and $F_{2}$ is in $\mathcal{V}%
(p_{1},q_{2})$, then the product $F_{1}F_{2}$ is in $\mathcal{V}%
(p_{1}+p_{2},q_{1}+q_{2})$.

The following Lemma provides the tool that we use to estimate the terms on
the right hand side of (\ref{repi}).

\begin{lemma}
\label{lem:estimations} \label{lem:bornes} Let $F\in\mathcal{V}(F_{0},F_{1}) 
$ and $G\in\mathcal{V}(G_{0},G_{1})$, and assume that 
\begin{align}
1-G_{1} & <F_{1}~,  \label{cb1} \\
1+G_{0} & >-F_{0}~,  \label{cb2}
\end{align}
and that 
\begin{equation}
F_{1}\neq1+G_{0}~.  \label{cb3}
\end{equation}
Then, there is a constant $C>0,$ such that for all $t\geq1,$ 
\begin{equation}
\int_{0}^{\infty}\left| G(x)F(t^{\gamma}x)\right| ~dx\leq Ct^{-\xi }~,
\label{decay}
\end{equation}
where 
\begin{equation}
\xi=\gamma\cdot\min\{F_{1},1+G_{0}\}~.  \label{cb4}
\end{equation}
\end{lemma}

\begin{proof}
From (\ref{cb3}) it follows that either $F_{1}<1+G_{0}$ or $F_{1}>1+G_{0}.$
In the first case we get using (\ref{cb1}) that $1-G_{1}<$ $F_{1}<$ $%
1+G_{0}, $ and therefore 
\begin{align*}
\int_{0}^{\infty}\left| G(x)F(t^{\gamma}x)\right| ~dx & \leq\left(
\sup_{x\geq0}x^{F_{1}}\left| F(t^{\gamma}x)\right| \right) \int_{0}^{\infty}%
\frac{1}{x^{F_{1}}}\left| G(x)\right| ~dx \\
& \leq\mathrm{const.}~t^{-\gamma F_{1}}~,
\end{align*}
and in the second case we get using (\ref{cb2}) that $-F_{0}<1+G_{0}<F_{1},$
and therefore 
\begin{align*}
\int_{0}^{\infty}\left| G(x)F(t^{\gamma}x)\right| ~dx & \leq\left(
\sup_{x\geq0}\frac{1}{x^{G_{0}}}\left| G(x)\right| \right)
\int_{0}^{\infty}x^{G_{0}}\left| F(t^{\gamma}x)\right| ~dx \\
& \leq\mathrm{const.}~t^{-\gamma(1+G_{0})}~.
\end{align*}
\end{proof}

We now show that the right hand side in (\ref{repi}) can be bounded by a sum
of terms of the form $t^{-\sigma}G(y)F(z).$ For each of these terms we then
show that the corresponding functions $G$ and $F$ satisfy the hypothesis of
Lemma \ref{lem:estimations}. This then implies that $\int_{0}^{\infty
}t^{-\sigma}G(x)F(t^{\gamma}x)~dx\leq\mathrm{const.}$ $t^{-(\sigma+\xi)},$
for a certain $\xi$ depending on $F$ and $G.$ It is therefore sufficient to
prove that $\sigma+\xi\geq1+4\gamma$ for all these terms in order to prove
the inequality (\ref{eqn:thminhomo}).

\begin{proposition}
For $y,$ $z>0$ we have the bounds 
\begin{align}
T_{2}(y,z) & \leq\widehat{T_{2}}(y)\equiv2\left| \mu_{1}-\kappa y\right|
y^{-\delta}\left( \sup_{z>0}\left| z^{\delta}\eta(z)\right| \right) +2\left|
\mu_{1}\mu_{3}\right| ~,  \notag \\
T_{3}(y,z) & \leq\widehat{T_{3}}(z)\equiv\left| \left( 2\kappa
z+2\eta\right) \varphi_{2}\right| +2\left( \sup_{y>0}\left| \mu
_{3}(y)y^{\delta-2}\right| \right) z^{2-\delta}\eta~,  \notag \\
T_{4}(y,z) & \leq T_{4,1}(y)\cdot T_{4,2}(z)\equiv\left( 2\left| \mu
_{1}(y)-\kappa y\right| y^{2-\delta^{\prime}}\right) \cdot\left(
z^{\delta^{\prime}-2}\left| \varphi_{2}(z)\right| \right) ~,
\label{inequalities} \\
T_{6}(y,z) & \leq\widehat{T_{6}}(z)\equiv2\left( \sup_{y>0}\left| \mu
_{3}(y)y^{\delta-2}\right| \right) z^{2-\delta}\left| \varphi_{2}\right|
+\left| \varphi_{2}\right| ^{2}~,  \notag
\end{align}
and $T_{1}\in\mathcal{V}(0,\delta-1)$, $\widehat{T_{2}}\in\mathcal{V}%
(3-\delta,\delta-1)$, $\widehat{T_{3}}\in\mathcal{V}(2-\delta,\delta^{\prime
}-3)$, $T_{4,1}\in\mathcal{V}(5-\delta^{\prime},\delta^{\prime}-3)$, $%
T_{4,2}\in\mathcal{V}(\delta^{\prime}-\delta,0)$, $T_{5}\in\mathcal{V}%
(4-2\delta,2\delta)$, $\widehat{T_{6}}\in\mathcal{V}(4-2\delta,\delta^{%
\prime }+\delta-4)$.
\end{proposition}

\begin{proof}
The inequalities (\ref{inequalities}) follow by using the triangle
inequality and the asymptotic properties of the functions $\mu_{1}$, $%
\mu_{2} $, $\eta$ and $\varphi_{1}$.
\end{proof}

\paragraph{Bound on the function $A_{2}$.}

We have the bound 
\begin{equation*}
|A_{2}|\leq\mathrm{const.}\sum_{p=1}^{n-1}t^{-\sigma}\widehat{T_{2}}%
^{p}~\left( T_{1}^{n-1-p}\widehat{T_{3}}\right) ~,
\end{equation*}
where $\sigma=1+\gamma+3p\varepsilon\gamma.$ The function $G=\widehat{T_{2}}%
^{p}$ is in $\mathcal{V}((3-\delta)p,3\varepsilon p)$, and the function $%
F=T_{1}^{n-1-p}\widehat{T_{3}}$ is in $\mathcal{V}(2-\delta,$ $3\varepsilon
(n-1-p)-3+\delta^{\prime})$. Since $\delta^{\prime}>1,$ the inequalities (%
\ref{cb1}) and (\ref{cb2}) are satisfied and, since $\delta^{\prime}<5$ for $%
n\geq3,$%
\begin{equation*}
\xi/\gamma=\left\{ 
\begin{array}{ll}
\delta^{\prime}-3\varepsilon p & \text{if}\;p\geq2 \\ 
3-3\varepsilon p & \text{if}\;p=1
\end{array}
\right. \geq3-3\varepsilon p~.
\end{equation*}
Therefore, $\sigma+\xi\geq$ $1+\gamma+3p\varepsilon\gamma+(3-3p\varepsilon
)\gamma=$ $1+4\gamma\ $as required.

\paragraph{Bound on the function $A_{3}$.}

We have that $A_{3}=t^{-\gamma(1+\delta^{\prime})}B_{3,4}+t^{-2\varepsilon
}B_{3,5}+t^{-6\gamma}B_{3,6},$ where $B_{3,i}=n(t^{-2\gamma}T_{1}+t^{-%
\varepsilon}T_{2})^{n-1}T_{i},$ $i=4,\dots,6$. Since $|T_{2}/T_{1}|\leq%
\mathrm{const.}~t^{3\gamma\varepsilon}$, and $\varepsilon-3\gamma
\varepsilon=2\gamma,$ we have the bound 
\begin{equation*}
t^{-\gamma(1+\delta^{\prime})}|B_{3,4}|\leq\mathrm{const.}\text{ }t^{-\sigma
}\left( T_{1}^{n-1}T_{4,2}\right) ~T_{4,1}~,
\end{equation*}
with $\sigma=2\gamma(n-1)+\gamma(1+\delta^{\prime}).$ The function $%
G=T_{4,1} $ is in $\mathcal{V}(5-\delta^{\prime},\delta^{\prime}-3)$ and the
function $F=T_{1}^{n-1}T_{4,2}$ is in $\mathcal{V}(\delta^{\prime}-\delta,3)$%
. Since $\delta^{\prime}>3$ the inequalities (\ref{cb1}) and (\ref{cb2}) are
satisfied and $\xi/\gamma=6-\delta^{\prime}$. Therefore $\sigma+\xi=$ $%
1-3\gamma +\gamma(1+\delta^{\prime})+(6-\delta^{\prime})\gamma=$ $1+4\gamma$
as required.

Similarly, we have that 
\begin{equation*}
t^{-2\varepsilon}|B_{3,5}|\leq\mathrm{const.}\text{ }t^{-%
\sigma}T_{1}^{n-1}~T_{5}~,
\end{equation*}
with $\sigma=2\gamma(n-1)+2\varepsilon.$ The function $G=T_{5}$ is in $%
\mathcal{V}(4-2\delta,2\delta)$ and the function $F=T_{1}^{n-1}$ is in $%
\mathcal{V}(0,3)$. The inequalities (\ref{cb1}) and (\ref{cb2}) are
satisfied and $\xi/\gamma=5-2\delta$ $=3-6\varepsilon.$ Therefore, $%
\sigma+\xi=$ $1-3\gamma+2\varepsilon+(3-6\varepsilon)\gamma=$ $1+4\gamma$ as
required.

Finally, 
\begin{equation*}
t^{-6\gamma}|B_{3,6}|\leq\mathrm{const.}\text{ }t^{-\sigma}T_{1}^{n-1}%
\widehat{T_{6}}~,
\end{equation*}
where $\sigma=2(n-1)\gamma+6\gamma.$ The function $G\equiv1$ is in $\mathcal{%
V}(0,0)$, and the function $F=T_{1}^{n-1}\widehat{T_{6}}$ is in $\mathcal{V}%
(4-2\delta,3(n-1)\varepsilon+\delta^{\prime}+\delta-4)$. The inequalities (%
\ref{cb1}) and (\ref{cb2}) are satisfied and $\xi/\gamma=1$. Therefore, $%
\sigma+\xi=$ $1-3\gamma+6\gamma+\gamma=$ $1+4\gamma$ as required.

\paragraph{Bound on the function $A_{4}$.}

Since the functions $T_{3}/T_{1}$ and $T_{6}/T_{1}$ are bounded, $%
T_{4}/T_{1}\leq\mathrm{const.}~t^{3\varepsilon\gamma}$ and $T_{5}/T_{1}\leq%
\mathrm{const.}~t^{3\varepsilon\gamma}$ we have that 
\begin{equation*}
|A_{4}|\leq\mathrm{const.}\sum_{p=2}^{n}t^{-2\gamma(n+p)}T_{1}^{n}\leq%
\mathrm{const.}~t^{-\sigma}T_{1}^{n}~,
\end{equation*}
where $\sigma=2n\gamma+4\gamma.$ The function $G\equiv1$ is in $\mathcal{V}%
(0,0)$, and the function $F=T_{1}^{n}$ is in $\mathcal{V}(0,3n\varepsilon) $%
. The inequalities (\ref{cb1}) and (\ref{cb2}) are satisfied and $%
\xi/\gamma=1$. Therefore, $\sigma+\xi=$ $\left( 1-\gamma\right)
+4\gamma+\gamma=$ $1+4\gamma$ as required.

\paragraph{Bound on the function $A_{5}$.}

We have the bound 
\begin{equation}
\left| A_{5}\right| \leq t^{-\sigma}\left| 3\gamma\varphi_{2}+\alpha
z\varphi_{2}^{\prime}\right| ~,
\end{equation}
where $\sigma=2n\gamma+4\gamma.$ The function $G\equiv1$ is in $\mathcal{V}%
(0,0)$, and the function $F=\left| 3\gamma\varphi_{2}+\alpha z\varphi
_{2}^{\prime}\right| $ is in $\mathcal{V}(2-\delta,\delta^{\prime}-2)$. The
inequalities (\ref{cb1}) and (\ref{cb2}) are satisfied and $\xi/\gamma=1$.
Therefore, $\sigma+\xi=$ $\left( 1-\gamma\right) +4\gamma+\gamma=$ $%
1+4\gamma $ as required.

\paragraph{Bound on the function $A_{6}$.}

We have the bound 
\begin{equation*}
|A_{6}|\leq\mathrm{const.}\text{ }t^{-\sigma}\left( T_{1}^{n-1}z^{2-\delta
}\eta\right) ~|\lambda_{0}-\mu_{3}y^{\delta-2}|~,
\end{equation*}
where $\sigma=1+\gamma.$ The function $G=|\lambda_{0}-\mu_{3}y^{\delta-2}|$
is in $\mathcal{V}(2,2)$ and the function $F=T_{1}^{n-1}z^{2-\delta}\eta$ is
in $\mathcal{V}(2-\delta,1+2\delta).$ The inequalities (\ref{cb1}) and (\ref
{cb2}) are satisfied and $\xi/\gamma=3$. Therefore, $\sigma+\xi=$ $%
1+\gamma+3\gamma=$ $1+4\gamma$ as required.

\paragraph{Bound on the function $A_{7}$.}

We have the bound 
\begin{equation*}
|A_{7}|\leq\mathrm{const.}\text{ }t^{-\sigma}|T_{1}^{n-1}-(2\kappa\lambda
z^{1-\delta})^{n-1}|~|2\mu_{1}\mu_{3}-2\kappa\lambda_{0}y^{3-\delta}|~,
\end{equation*}
where $\sigma=2n\gamma-2\gamma+\varepsilon$. The function $G=|2\mu_{1}\mu
_{3}-2\kappa\lambda_{0}y^{3-\delta}|$ is in $\mathcal{V}(5-\delta,\delta-3)$
and the function $F=|T_{1}^{n-1}-(2\kappa\lambda z^{1-\delta})^{n-1}|$ is in 
$\mathcal{V}(-3,3+\delta^{\prime}-\delta).$ The inequalities (\ref{cb1}) and
(\ref{cb2}) are satisfied and $\xi/\gamma=6-\delta$. Therefore, $\sigma+\xi=$
$\left( 1-\gamma\right) -2\gamma+\varepsilon+(6-\delta)\gamma=$ $1+4\gamma$
as required.

\paragraph{Bound on the function $A_{8}$.}

We have the bound 
\begin{equation*}
|A_{8}|\leq \mathrm{const.}\text{ }t^{-\sigma }|T_{1}^{n-1}\eta -(2\kappa
\lambda z^{1-\delta })^{n-1}\lambda z^{-\delta }|~|\mu _{1}-\kappa y-\kappa
_{3}y^{3}|~,
\end{equation*}
where $\sigma =2n\gamma -\gamma .$ The function $G=|\mu _{1}-\kappa y-\kappa
_{3}y^{3}|$ is in $\mathcal{V}(5,-3)$ and the function $F=|T_{1}^{n-1}\eta
-(2\kappa \lambda z^{1-\delta })^{n-1}\lambda z^{-\delta }|\ $is in $%
\mathcal{V}(-3-\delta ,3+\delta ^{\prime }).$ The inequalities (\ref{cb1})
and (\ref{cb2}) are satisfied and $\xi /\gamma =6$. Therefore, $\sigma +\xi
= $ $\left( 1-\gamma \right) -\gamma +6\gamma =$ $1+4\gamma $ as required.

\paragraph{Bound on the functions $A_{p,q}$.}

We have the bound 
\begin{equation*}
|A_{p,q}|\leq\mathrm{const.}~t^{-\sigma}~|(z^{\delta}%
\eta)^{p-q}~T_{1}^{n-p}-\lambda^{p-q}(2\kappa\lambda
z^{1-\delta})^{n-p}|~\left( \left| 2(\mu_{1}-\kappa y)y^{-\delta}\right|
^{p-q}\left| 2\mu_{1}\mu_{3}\right| ^{q}\right) ~,
\end{equation*}
where $\sigma=2n\gamma-2p\gamma+p\varepsilon.$ The function $G=\left|
2(\mu_{1}-\kappa y)y^{-\delta}\right| ^{p-q}\left| 2\mu_{1}\mu_{3}\right|
^{q}$ is in $\mathcal{V}(p(3-\delta),3\varepsilon p+q)$ and the function $%
F=|(z^{\delta}\eta)^{p-q}~T_{1}^{n-p}-\lambda^{p-q}(2\kappa z^{1-\delta
})^{n-p}|$ is in $\mathcal{V}(-3\varepsilon(n-p),2+\delta^{\prime
}-3p\varepsilon)$. The inequalities (\ref{cb1}) and (\ref{cb2}) are
satisfied, and 
\begin{equation*}
\xi/\gamma=\left\{ 
\begin{array}{ll}
5-3p\varepsilon & \text{if}\;p=2~, \\ 
2+\delta^{\prime}-3p\varepsilon & \text{if}\;p\geq3~.
\end{array}
\right.
\end{equation*}
Therefore, $\sigma+\xi=1+4\gamma$, for $p=2$ and $\sigma+\xi=1+\gamma
(1+\delta^{\prime})>1+4\gamma$, for $p\geq3,$ as required.

\bigskip

\noindent This completes the proof of Theorem \ref{thm:Inhomo}. $%
\blacksquare $

\subsection{The Map $T$}

Equation (\ref{equpsi1}) is of the form (\ref{eqpsi}) if we define the map $%
T $ by the equation 
\begin{equation}
T(\psi )=\left\{ 
\begin{array}{ll}
\widehat{T}(\psi ) & \text{for }n\text{ odd}~, \\ 
\widehat{T}(\psi )+V_{1}\psi ~ & \text{for }n\text{ even}~,
\end{array}
\right.  \label{mapt}
\end{equation}
with $\widehat{T}$ as defined in (\ref{that}) and $V_{1}$ as defined in
Section \ref{sectionv}. Using the definitions, we see that $T$ can be
written as, 
\begin{equation}
T(\psi )=\sum_{p=1}^{n}\sum_{q=0}^{p}V_{p,q}~\psi ^{p+q}~,
\label{eqn:tpsidef}
\end{equation}
with 
\begin{equation}
V_{p,q}=\left\{ 
\begin{array}{ll}
0 & \text{for }(p,q)=(1,0)\text{ and }n\text{ odd,} \\ 
V_{1} & \text{for }(p,q)=(1,0)\text{ and }n\text{ even,} \\ 
\binom{n}{p}\binom{p}{q}(2\overline{u}\phi +\phi ^{2})^{n-p}(2\overline{u}%
+2\phi )^{p-q} & \text{for }p+q\geq 2~.
\end{array}
\right.  \label{vpq}
\end{equation}

\begin{proposition}
Let $V_{p,q}$ as in (\ref{vpq}). Then, for all $t\geq1,$ 
\begin{equation}
\sup_{x\in\mathbf{R}}|V_{p,q}(x,t)|\leq\mathrm{const.}\text{ }t^{-e(p,q)}~,
\label{eqn:Lemmavmp}
\end{equation}
where 
\begin{equation}
e(p,q)=\left\{ 
\begin{array}{ll}
\gamma(n-1)(\delta^{\prime}+1) & \text{for }(p,q)=\left( 1,0\right) ~, \\ 
2\gamma(n-2)+2\gamma & \text{for}\;(p,q)=(2,0)~, \\ 
2\gamma(n-p) & \text{for}\;(p,q)\neq(2,0)\;\text{and}\;p+q\geq2~.
\end{array}
\right.  \label{mupq}
\end{equation}
\end{proposition}

\begin{proof}
The case $(p,q)=(1,0)$ follows from (\ref{v1bound}). Let now $(p,q)\neq
(1,0).$ Since $\varepsilon -\gamma -\nu \gamma \geq 0,$ for all $\nu ,$ $%
0\leq \nu \leq \delta ,$ we find that 
\begin{align*}
\sup_{x\in \mathbf{R}}\left| z\right| ^{\nu }~\left| \phi (x,t)\right| &
\leq t^{-\gamma }(\sup_{z\in \mathbf{R}_{+}}\left| z^{\nu }~\eta (z)\right|
+t^{-(\varepsilon -\gamma -\nu \gamma )}\sup_{y\in \mathbf{R}_{+}}\left|
y^{\nu }~\mu _{3}(y)\right| +t^{-2\gamma }\sup_{z\in \mathbf{R}_{+}}\left|
z^{\nu }~\varphi _{2}(z)\right| ) \\
& \leq \mathrm{const.}\text{ }t^{-\gamma }~.
\end{align*}
Furthermore, since $\mu _{1}(y)=\mathcal{O}(y)$ near $y=0,$%
\begin{equation*}
\left| 2\overline{u}\phi +\phi ^{2}\right| \leq t^{-\gamma }~2\frac{\mu
_{1}(\left| y\right| )}{\left| y\right| }~\left| z\phi \right| +\left| \phi
\right| ^{2}\leq \mathrm{const.}\text{ }t^{-2\gamma }~.
\end{equation*}
Since the function $\left| \overline{u}+\phi \right| $ is bounded, it
follows that $|V_{p,q}(x,t)|\leq \mathrm{const.}$ $t^{-e(p,q)},$ with $%
e(p,q)=2\gamma (n-p).$ For $(p,q)=(2,0)$ we improve this bound using
additional properties of the function $\overline{u}+\phi .$ Namely, since $%
2/(n-2)\leq \delta -1,$ we have that 
\begin{align*}
|V_{2,0}(x,t)|& \leq \mathrm{const.}\sup_{x\in \mathbf{R}}\left| (2\overline{%
u}\phi +\phi ^{2})^{n-2}(\overline{u}+\phi )^{2}\right| \\
& \leq \mathrm{const.}\sup_{x\in \mathbf{R}}\left| (2t^{-\gamma }\frac{%
\overline{u}}{y}z\phi +\phi ^{2})^{n-2}(t^{-\gamma }z\frac{\overline{u}}{y}%
+\phi )^{2}\right| \\
& \leq \mathrm{const.}~\sup_{x\in \mathbf{R}}\left| 2t^{-\gamma }\frac{%
\overline{u}}{y}z^{1+2/(n-2)}\phi +\left( z^{2/(n-2)}\phi \right) ~\phi
\right| ^{n-2}~\left| t^{-\gamma }\frac{\overline{u}}{y}\right| ^{2} \\
& \hspace{3cm}+\mathrm{const.}~\sup_{x\in \mathbf{R}}\left| 2t^{-\gamma }%
\frac{\overline{u}}{y}z\phi +\phi ^{2}\right| ^{n-2}~\left( 2t^{-\gamma
}\left| \frac{\overline{u}}{y}\right| \left| z\phi \right| +\left| \phi
\right| ^{2}\right) \\
& \leq \mathrm{const.}~t^{-2(n-2)\gamma -2\gamma }~.
\end{align*}
\end{proof}

\section{Proof of the main result}

For functions $f$ in $\mathcal{J}=L_{1}(\mathbf{R)}\cap L_{\infty }(\mathbf{%
R)}$ we use the norms $\left\| f\right\| _{1}=\int |f(x)|~dx$, $\left\|
f\right\| _{\infty }=\sup_{x\in \mathbf{R}}|f(x)|$ and $\left\| f\right\|
=\left\| f\right\| _{1}+\left\| f\right\| _{\infty }$, and we denote by $%
\mathcal{B}$ the Banach space of functions $\varphi $ in $L_{\infty
}([1,\infty ))\times \mathcal{J}$ for which the norm $\left\| ~~\right\| _{%
\mathcal{B}}$, 
\begin{equation*}
\left\| \varphi \right\| _{\mathcal{B}}=\sup_{t\geq 1}t^{4\gamma }\left\|
\varphi (\sqrt{t}~.~,t)\right\| ~,
\end{equation*}
is finite. Let $\tau _{0}$ as in Proposition \ref{propvodd} and $\tau _{1}$
as in Proposition \ref{propV}, and consider, for fixed $\tau >\max \{\tau
_{0},\tau _{1}\},$ functions $\psi $ of the form 
\begin{equation*}
\psi (x,t)=\tau ^{-4\gamma }\varphi (x/\sqrt{\tau },t/\tau )~,
\end{equation*}
with $\varphi \in \mathcal{B}.$ Let $K$ be the fundamental solution of the
differential operator $\partial _{t}-\partial _{x}^{2}-\tau V(\sqrt{\tau }%
x,\tau t),$ and let, for given $\nu \in \mathcal{J}$, the map $\mathcal{R}$
be defined by the equation 
\begin{equation*}
\mathcal{R}(\varphi )(x,t)=\varphi _{0,1}(x,t)+\varphi _{0,2}(x,t)+\mathcal{N%
}(\varphi )(x,t)~,
\end{equation*}
where 
\begin{align*}
\varphi _{0,1}(x,t)& =\int_{\mathbf{R}}K(x,t;y,1)~\nu (y)~dy~, \\
\varphi _{0,2}(x,t)& =\tau ^{4\gamma }\tau \int_{1}^{t}ds\int_{\mathbf{R}%
}dy~K(x,t;y,s)~I(\sqrt{\tau }y,\tau s)~,
\end{align*}
and where 
\begin{equation*}
\mathcal{N}(\varphi )(x,t)=\sum_{p=1}^{n}\sum_{q=0}^{p}\mathcal{N}%
_{p,q}(\varphi )(x,t)~,
\end{equation*}
with 
\begin{equation*}
\mathcal{N}_{p,q}(\varphi )(x,t)=\tau ^{4\gamma }\tau \int_{1}^{t}ds\int_{%
\mathbf{R}}dy~K(x,t;y,s)~V_{p,q}(\sqrt{\tau }y,\tau s)~\tau ^{-4\gamma
(p+q)}\varphi (y,s)^{p+q}~.
\end{equation*}
The integral equation $\varphi =\mathcal{R}(\varphi )$ is equivalent to the
differential equation (\ref{eqpsi}) with initial condition $\psi
_{0}(x)=\psi (x,\tau )=\tau ^{-4\gamma }\nu (x/\sqrt{\tau }).$ We note that,
since the function $V$ is positive, the kernel $K$ is bounded pointwise by
the fundamental solution $K_{0}$ of the heat equation, 
\begin{equation}
K_{0}(x,t;y,s)=\frac{1}{\sqrt{4\pi }}\frac{1}{\sqrt{t-s}}\exp \left( -\frac{1%
}{4}\frac{(x-y)^{2}}{(t-s)}\right) ~.
\end{equation}
The following Proposition makes Theorem \ref{main} precise.

\begin{proposition}
\label{maint}Let $\beta \geq \max \left\{ 1,~3~c_{I}\int_{0}^{1}\left( 1+%
\frac{1}{\sqrt{1-s}}\right) \frac{ds}{s^{1/2+4\gamma }}\right\} ,$ with $%
c_{I}$ as defined in (\ref{eqn:thminhomo}), and let $\tau $ be sufficiently
large. Then, for all $\nu \in \mathcal{J}$ with $\left\| \nu \right\| <\beta
/6$, the equation $\varphi =\mathcal{R}(\varphi )$ has a unique solution $%
\varphi ^{\ast }$ in the ball $\mathcal{U}(\beta )=\{\varphi \in \mathcal{B}%
| $ $\left\| \varphi \right\| _{\mathcal{B}}<\beta \}.$
\end{proposition}

\begin{proof}
Since $4\gamma <1/2,$ the solution of the integral equation will be
dominated by $\varphi _{0,2},$ and, as we will see, $\beta $ has been chosen
such that $\left\| \varphi _{0,2}\right\| _{\mathcal{B}}\leq \beta /3.$ The
idea is therefore to show that, if $\tau $ is large enough to make the
nonlinear part of the map $\mathcal{R}$ small, and if $\left\| \nu \right\|
<\beta /6$, then the map $\mathcal{R}$ contracts the ball $\mathcal{U}(\beta
)$ into itself, which by the contraction mapping principle implies the
theorem. We first show that $\mathcal{R}$ maps the ball $\mathcal{U}(\beta )$
into itself. For the contribution coming from the initial condition we have 
\begin{equation*}
\left\| \varphi _{0,1}(\sqrt{t}~.~,t)\right\| \leq \frac{2}{\sqrt{t}}\left\|
\nu \right\| ~,
\end{equation*}
and therefore 
\begin{equation*}
\left\| \varphi _{0,1}\right\| _{\mathcal{B}}\leq 2\left\| \nu \right\|
<\beta /3~.
\end{equation*}
We next estimate the norm $\left\| \varphi _{0,2}\right\| _{\mathcal{B}}.$
Let $c(t,s)=\frac{1}{\sqrt{t}}+\frac{1}{\sqrt{t-s}}.$ Then, 
\begin{align*}
\left\| \varphi _{0,2}(\sqrt{t}~.~,t)\right\| & \leq \tau ^{4\gamma }\tau
\int_{1}^{t}ds~c(t,s)\int_{\mathbf{R}}dy~\left| I(\sqrt{\tau }y,\tau
s)\right| \\
& =\tau ^{4\gamma }\tau \int_{1}^{t}\sqrt{s}~c(t,s)~ds~\int_{\mathbf{R}%
}dx~\left| I(\sqrt{\tau s}x,\tau s)\right| \\
& \leq c_{I}~\tau ^{4\gamma }\tau \int_{1}^{t}\sqrt{s}~c(t,s)~ds~\left( \tau
s\right) ^{-(1+4\gamma )} \\
& \leq c_{I}~t^{-4\gamma }\int_{0}^{1}c(1,s)~\frac{ds}{s^{1/2+4\gamma }} \\
& \leq \frac{\beta }{3}t^{-4\gamma }~,
\end{align*}
and therefore 
\begin{equation*}
\left\| \varphi _{0,2}\right\| _{\mathcal{B}}<\beta /3~.
\end{equation*}
It remains to be shown that the nonlinearity is also bounded by $\beta /3,$
for $\tau $ large enough. For $\varphi \in \mathcal{U}(\beta )$ we have, 
\begin{equation}
\left\| \mathcal{N}(\varphi )(\sqrt{t}~.~,t)\right\| \leq \mathrm{const.}%
~\tau ^{4\gamma }\tau \int_{1}^{t}c(t,s)~\sqrt{s}~ds\sum_{p=1}^{n}%
\sum_{q=0}^{p}(\tau s)^{-e(p,q)}~s^{-4\gamma (p+q)}\tau ^{-4\gamma
(p+q)}\left\| \varphi \right\| _{\mathcal{B}}^{p+q}~.  \label{normcase}
\end{equation}
For $(p,q)=(1,0)$ we get, since $\delta _{1}\equiv \gamma (n-1)(\delta
^{\prime }+1)-1>0,$ 
\begin{align*}
\left\| \mathcal{N}_{1,0}(\varphi )(\sqrt{t}~.~,t)\right\| & \leq \mathrm{%
const.}~\tau ^{-\delta _{1}}\beta \int_{1}^{t}c(t,s)~\sqrt{s}%
~ds~s^{-1-\delta _{1}-4\gamma } \\
& \leq \mathrm{const.}~\tau ^{-\delta _{1}}\beta ~t^{-4\gamma }\int_{0}^{1}%
\frac{c(1,s)}{s^{1/2+4\gamma }}~ds~,
\end{align*}
and for $(p,q)=(2,0)$ we get 
\begin{align*}
\left\| \mathcal{N}_{2,0}(\varphi )(\sqrt{t}~.~,t)\right\| & \leq \mathrm{%
const.}~\tau ^{4\gamma +1-8\gamma -(2\gamma (n-2)+2\gamma )}\beta
^{2}\int_{1}^{t}c(t,s)~\sqrt{s}~ds~s^{-8\gamma -2\gamma -2\gamma (n-2)} \\
& \leq \mathrm{const.}~\tau ^{-\gamma }\beta ^{2}~t^{-4\gamma }\int_{0}^{1}%
\frac{c(1,s)}{s^{1/2+4\gamma }}~ds~,
\end{align*}
and for the other cases we have 
\begin{align*}
\left\| \mathcal{N}_{p,q}(\varphi )(\sqrt{t}~.~,t)\right\| & \leq \mathrm{%
const.}~\tau ^{4\gamma +1-2\gamma (n-p)-4\gamma p-4\gamma q}\beta
^{2n}\int_{1}^{t}c(t,s)~\sqrt{s}~ds~s^{-2\gamma (n-p)-4\gamma p-4\gamma q} \\
& \leq \mathrm{const.}~\tau ^{-\gamma }\beta ^{2n}~t^{-4\gamma }\int_{0}^{1}%
\frac{c(1,s)}{s^{1/2+4\gamma }}~ds~,
\end{align*}
and therefore $\left\| \mathcal{N}(\varphi )\right\| _{\mathcal{B}}\leq
\beta /3$ if $\tau $ is large enough. Using the triangle inequality we get
that $\left\| \mathcal{R}(\varphi )\right\| _{\mathcal{B}}\leq \beta ,$
which proves that $\mathcal{R}\left( \mathcal{U}(\beta )\right) \subset 
\mathcal{U}(\beta )$ as claimed. We now show that $\mathcal{R}$ is
Lipschitz. Let $\varphi _{1}$ and $\varphi _{2}$ be in $\mathcal{U}(\beta ).$
We have 
\begin{align*}
\left\| \mathcal{N}(\varphi _{1})(\sqrt{t}~.~,t)-\mathcal{N}(\varphi _{2})(%
\sqrt{t}~.~,t)\right\| & \leq \mathrm{const.}~\tau ^{4\gamma }\tau
\int_{1}^{t}ds~c(t,s)~\sqrt{s}\cdot \\
& \sum_{p=1}^{n}\sum_{q=0}^{p}(\tau s)^{-e(p,q)}~s^{-4\gamma (p+q)}\tau
^{-4\gamma (p+q)}~\beta ^{p+q-1}~\left\| \varphi _{1}-\varphi _{2}\right\| _{%
\mathcal{B}}~,
\end{align*}
and therefore we get, using the same estimates as for (\ref{normcase}), that 
\begin{equation*}
\left\| \mathcal{R}(\varphi _{1})-\mathcal{R}(\varphi _{2})\right\| _{%
\mathcal{B}}=\left\| \mathcal{N}(\varphi _{1})-\mathcal{N}(\varphi
_{2})\right\| _{\mathcal{B}}\leq \frac{1}{2}\left\| \varphi _{1}-\varphi
_{2}\right\| _{\mathcal{B}}~,
\end{equation*}
provided $\tau $ is large enough. This completes the proof of Theorem \ref
{maint}.
\end{proof}

\section{Appendix}

\subsection{Proof of Proposition \ref{theoremeta}}

We first prove the existence of a unique positive solution of equation (\ref
{eqeta}) satisfying the boundary conditions (\ref{etainfinity}) and (\ref
{etazero}). Then, we derive the results on the asymptotic behavior near zero
and infinity.

\subsubsection{Existence of the function $\protect\eta\label{seta777}$}

\begin{proposition}
Let, for $\rho >0$, $\eta _{\rho }$ be the solution of the initial value
problem on $\mathbf{R}_{+}$, 
\begin{align}
\eta ^{\prime \prime }& =(2\kappa z\eta +\eta ^{2})^{n}~,  \label{1} \\
\eta ^{\prime }(0)& =-\kappa ~,  \notag \\
\eta (0)& =\rho >0~.  \notag
\end{align}
Then, there exists a unique $\bar{\rho}$ such that the function $\eta _{%
\overline{\rho }}$ is positive and satisfies $\lim_{x\rightarrow \infty
}\eta _{\bar{\rho}}(x)=0$.
\end{proposition}

\begin{proof}
We first prove that $\overline{\rho}$ is unique. Given a function $\eta$
from $\mathbf{R}_{+}$ to $\mathbf{R}$ we define the function $\mathcal{F}%
(\eta),$ $\mathcal{F}\left( \eta\right) (z)=(\kappa z\eta+\eta^{2})^{n}$.
Assume that there are two values $\rho_{1}>\rho_{2}>0,$ such that the
functions $\eta _{1}\equiv\eta_{\rho_{1}}$ and $\eta_{2}\equiv\eta_{%
\rho_{2}} $ are positive and satisfy $\lim_{x\rightarrow\infty}\eta_{1}(x)=$ 
$\lim_{x\rightarrow\infty }\eta_{2}(x)=0.$ We first show that the function $%
\eta_{12}=\eta_{1}-\eta_{2}$ is positive for all $x\geq0.$ Namely, if we
assume the contrary, then because $\eta_{12}(0)>0$, there must be a first $%
x_{0}>0$ such that $\eta_{12}(x_{0})=0.$ Furthermore, if $\eta_{12}(x)>0$
then $\eta_{12}^{\prime\prime }(x)=\mathcal{F}\left( \eta_{1}\right) (x)-%
\mathcal{F}\left( \eta _{2}\right) (x)>0,$ and therefore $%
\eta_{12}(x_{0})=\rho_{1}-\rho_{2}+\int_{0}^{x_{0}}dx\int_{0}^{x}dy~%
\eta_{12}^{\prime\prime}(y)>0,$ a contradiction. Therefore $\eta_{12},$ and
as a consequence $\eta_{12}^{\prime\prime},$ are positive for all $x,$ from
which it follows that $\lim_{x\rightarrow\infty}\eta_{12}(x)>0,$ in
contradiction with $\lim_{x\rightarrow\infty}\eta_{12}(x)=\lim_{x\rightarrow%
\infty}\eta _{1}(x)-\lim_{x\rightarrow\infty}\eta_{2}(x)=0.$

To prove the existence of a $\bar{\rho}$ for which $\eta _{\overline{\rho }}$
is positive and for which $\lim_{x\rightarrow \infty }\eta _{\bar{\rho}%
}(x)=0 $, we use the so called shooting method. Note that, for any $\rho >0$%
, the initial value problem (\ref{1}) has a unique solution $\eta _{\rho },$
and since $\eta _{\rho }^{\prime }(0)=-\kappa ,$ the function $\eta _{\rho }$
is strictly decreasing on $[0,x_{\rho })$ for $x_{\rho }$ small enough. We
will show that for small enough $\rho >0$, the graph of $\eta _{\rho }$
intersects the real axis and $\eta _{\rho }$ becomes negative, whereas for $%
\rho $ large enough, $\eta _{\rho }$ has a minimum and then diverges to plus
infinity. The (unique) point between those two sets is $\bar{\rho}$. Define
the two subsets $I_{1}$ and $I_{2}$ of $\mathbf{R}_{+},$ 
\begin{align*}
I_{1}& =\{\rho \in \mathbf{R}_{+}|~\exists ~x_{1},\eta _{\rho }(x_{1})=0%
\text{{\ \textrm{and\ }}}\eta _{\rho }(x)>0{\ \mathrm{for\ }}x\in \lbrack
0,x_{1})\}~, \\
I_{2}& =\{\rho \in \mathbf{R}_{+}|~\exists ~x_{2},\eta _{\rho }^{\prime
}(x_{2})=0{\ \mathrm{and\ }}\eta _{\rho }^{\prime }(x)<0,\eta _{\rho }(x)>0{%
\ \mathrm{for\ }}x\in \lbrack 0,x_{2})\}~.
\end{align*}
We note that if $\eta _{\rho }^{\prime }(x_{0})=0$ and $\eta _{\rho
}(x_{0})>0,$ for some $x_{0}$, then $\eta _{\rho }^{\prime }>0$ on any
interval $(x_{0},x)$ on which $\eta _{\rho }$ is defined, and a function $%
\eta _{\rho }$ with $\rho \in I_{2}$ can therefore not converge to zero at
infinity. Furthermore, since the function $\eta \equiv 0$ is a solution of
the differential equation (\ref{1}), it follows, since solutions are unique,
that $\eta _{\rho }(x_{0})>0$ if $\eta _{\rho }^{\prime }(x_{0})=0,$ and
therefore the intersection of $I_{1}$ with $I_{2}$ is empty. The sets $I_{1}$
and $I_{2}$ are open, by continuity of the solution $\eta _{\rho }$ as a
function of the initial data $\rho $. We now show that $I_{1}$ is non empty
and bounded, which shows that $\bar{\rho}\equiv \sup I_{1}<\infty .$ This $%
\bar{\rho}$ is neither in $I_{1}$ nor in $I_{2}$, and therefore the function 
$\eta _{\bar{\rho}}$ is at the same time strictly positive and strictly
decreasing, and therefore $\lim_{x\rightarrow \infty }\eta _{\bar{\rho}%
}(x)=0 $. To prove that $I_{1}$ is non empty, we fix any $\rho _{1}$
positive and choose $x_{0}>0$ small enough such that on $[0,x_{0}]$ the
solution $\eta _{1}\equiv \eta _{\rho _{1}}$ exists and is strictly
decreasing. Then, $\rho _{1}-\eta _{1}(x_{0})>0.$ Choose now $0<\rho
_{2}<\rho _{1}-\eta _{1}(x_{0})$ and let $\eta _{2}\equiv \eta _{\rho _{2}}$
be the corresponding solution. As before, we have that the function $\eta
_{12}=\eta _{1}-\eta _{2},$ and its second derivative $\eta _{12}^{\prime
\prime },$ are positive on the interval $[0,x_{0}),$ and therefore, since $%
\eta _{2}(x_{0})=$ $\rho _{2}+\int_{0}^{x_{0}}dx\int_{0}^{x}dy~\eta
_{2}^{\prime \prime }(y)$ $=\rho _{2}+\eta _{1}(x_{0})-\rho
_{1}-\int_{0}^{x_{0}}dx\int_{0}^{x}dy~\eta _{12}^{\prime \prime }(y),$ we
find that $\eta _{2}(x_{0})<\rho _{2}-\rho _{1}+\eta _{1}(x_{0}).$ Using the
definition of $\rho _{2}$ we therefore find that $\eta _{2}(x_{0})<0.$
Therefore $\rho _{2}\in I_{1}$. We now prove that $I_{1}$ is bounded. For $%
\rho >0$, let $x_{\rho }$ be the largest value (possibly infinite) such that
on $[0,x_{\rho })$ the solution $\eta _{\rho }$ exists and is strictly
positive. Then, $\eta _{\rho }^{\prime \prime }=\mathcal{F}\left( \eta
_{\rho }\right) $ is positive on $(0,x_{\rho })$ and, therefore $\eta _{\rho
}(x)>\rho -\kappa x$ for $x\in (0,x_{\rho })$. As a consequence, if the
function $\eta _{\rho }$ exists on $[0,\rho /\kappa ],$ then $x_{\rho }\geq
\rho /\kappa .$ Using again that $\eta _{\rho }(x)>\rho -\kappa x$ we then
find that $\eta _{\rho }(x)>\rho /2$ for $x\in \lbrack 0,\rho /2\kappa ]$,
and therefore $\mathcal{F}\left( \eta _{\rho }\right) >(\rho /2)^{2n}$ on $%
[0,\rho /2\kappa ]$, which implies that $\eta _{\rho }^{\prime }(\rho
/2)>-\kappa +(\rho /2)^{2n+1},$ which is positive if $\rho >2\kappa
^{1/2n+1}.$ Therefore $\eta ^{\prime }(x)$ must be equal to zero for some $%
x<\rho /\kappa .$ Any such $\rho $ therefore belongs to $I_{2}.$ If the
function $\eta _{\rho }$ ceases to exist before $x=\rho /\kappa $ it must
have been diverging to plus infinity for some $x<\rho /\kappa $ which again
implies that $\eta _{\rho }^{\prime }(x)$ must have been equal to zero for
some $x<\rho /\kappa ,$ and the corresponding $\rho $ is in $I_{2}.$
\end{proof}

\subsubsection{Asymptotic behavior of the function $\protect\eta$}

The function $\eta $ is regular at zero, and the coefficients of its Taylor
series at zero can be computed recursively. We have $\eta (0)=\eta _{0}>0$
and $\eta ^{\prime }(0)=-\kappa ,$ and therefore we get using the
differential equation that $\eta _{2}=\eta ^{\prime \prime }(0)/2=\eta
_{0}^{2n}/2,$ $\eta ^{\prime \prime \prime }(0)=0$ and $\eta _{4}=-\eta
^{iv}(0)/4!=\frac{n}{12}\eta _{0}^{2n-2}(\kappa ^{2}-\eta _{0}^{2n+1}).$ The
asymptotic behavior of $\eta $ at infinity is obtained as follows. Assuming
that $\eta $ behaves like $\lambda /z^{\delta }$ at infinity we get from the
differential equation that $\delta $ and $\lambda $ are as defined in (\ref
{defdelta007}) and (\ref{deflambda007}), respectively. That this is indeed
the correct leading behavior of $\eta $ at infinity can now be proved by
using standard techniques based on repeated applications of l'H\^{o}pital's
rule. See for example \cite{Brezis}. Since the proof is simple, but lengthy
and quite uninteresting, we do not give the details here.

Once the leading behavior of $\eta $ at infinity has been established we
make the ansatz $\eta (z)=\lambda z^{-\delta }+s(z)$. To leading order we
get for the function $s$ the linear equation 
\begin{equation}
s^{\prime \prime }-\frac{n}{\lambda }(2\kappa \lambda )^{n}z^{-2}s=n(2\kappa
\lambda )^{n-1}\lambda ^{2}z^{-3-2\delta }~.  \label{eqs}
\end{equation}
There is a certain constant $\lambda _{p},$ such that the function $s_{p},$ $%
s_{p}(z)=\lambda _{p}z^{-1-2\delta }$ is a particular solution of equation (%
\ref{eqs}). The solutions of the homogeneous equation associated with (\ref
{eqs}) are of the form $s_{h}^{\pm }(z)=z^{p_{\pm }}$, and using the
definition (\ref{deflambda007}) for $\lambda $ we find that 
\begin{equation}
p_{\pm }=\frac{1}{2}\left( 1\pm \sqrt{1+4n\delta (\delta +1)}\right) ~.
\label{eqn:critexp}
\end{equation}
For $n\geq 6$ we have that $\left| p_{-}\right| >2\delta +1,$ and the
asymptotic behavior of $s$ is therefore for $n\geq 6$ of the form $\lambda
_{\infty }/z^{2\delta +1},$ with $\lambda _{\infty }=\lambda _{p}$, and of
the form $\lambda _{\infty }/z^{\left| p_{-}\right| }$ with some unknown
coefficient $\lambda _{\infty }$ for $n\leq 5$. It is tedious, but not
difficult, to prove that this is indeed the correct second order behavior of 
$\eta $ at infinity. We omit the details.

\subsection{Proof of Proposition \ref{cmu2}}

In order to study equation (\ref{mu2}) with boundary conditions (\ref{bmu2-1}%
) and (\ref{bmu2-2}), we make the ansatz $\mu_{2}(x)=m(x)/x^{\delta}.$ For
the function $m$ we get the differential equation 
\begin{equation}
m^{\prime\prime}+(\frac{x}{2}-\frac{2\delta}{x})m^{\prime}+\left(
\delta(\delta+1)\frac{1}{x^{2}}-\left( \frac{\delta}{2}-\varepsilon\right)
\right) m=\frac{1}{x^{2}}(2\frac{\mu_{1}(x)}{x}m)^{n}~,  \label{eqh}
\end{equation}
and the boundary conditions for $m$ are 
\begin{align}
\lim_{x\rightarrow0}m(x) & =\lambda~,  \label{bcm0} \\
\lim_{x\rightarrow\infty}m(x)x^{2\varepsilon-\delta} & =0~.  \label{bcm1}
\end{align}

\subsubsection{Asymptotic behavior of the function $\protect\mu_{2}$}

As indicated in Section \ref{smu2}, a solution of equation (\ref{mu2}) that
is defined on $\mathbf{R}_{+}$ behaves at infinity either like $%
x^{-2\varepsilon }$ or like $\exp (-x^{2}/4)/x^{1-2\varepsilon }.$ The proof
is similar to the one in \cite{Brezis}. We omit the details. Given the
asymptotic behavior of $\mu _{2}$ at infinity, we find for the function $m$
at infinity either a behavior proportional to $x^{\delta -2\varepsilon }$,
or a behavior proportional to $x^{5\varepsilon }\exp (-x^{2}/4).$ Since $%
\delta -2\varepsilon >0,$ we find that 
\begin{equation}
\lim_{x\rightarrow \infty }m(x)=0~,  \label{bcm3}
\end{equation}
if and only if the boundary condition (\ref{bcm1}) is satisfied, and we will
impose (\ref{bcm3}) from now on. We now discuss the asymptotic behavior of
the function $m$ near zero. From equation (\ref{eqh}) we see that $m^{\prime
\prime }(0)$ exists if and only if $\delta (\delta +1)m(0)=\left( \kappa
m(0)\right) ^{n}$, i.e., if $m(0)=\lambda ,$ and if $m^{\prime }(0)=0.$ We
then find, that $m^{\prime \prime }(0)/2=\lambda _{0}$, with $\lambda _{0}$
as defined in (\ref{lambda0}). By taking derivatives of equation (\ref{eqh})
we find that $m^{\prime \prime \prime }(0)=0$, and that $m^{iv}(0)/4!=%
\lambda _{1},$ for some constant $\lambda _{1}\neq 0.$ By taking further
derivatives, one can recursively compute the Taylor coefficients of a
solution $m_{0}$ of equation (\ref{eqh}) that is regular (in fact, analytic)
in a neighborhood of zero. The solution $m_{0}$ does however not satisfy the
boundary condition (\ref{bcm3}). The solution of (\ref{eqh}) that does
satisfy (\ref{bcm3}) is of the form 
\begin{equation}
m(x)=m_{0}(x)+x^{p}m_{1}(x)~,  \label{msing}
\end{equation}
where $p=p_{+}+\delta ,$ with $p_{+}$ as defined in (\ref{eqn:critexp}).
Here, $m_{1}(x)=m_{1}(0)+\dots $, with $m_{1}^{\prime }(0)=0,$ and with $%
m_{1}(0)$ to be determined. The asymptotic form (\ref{msing}) can be
obtained by substituting the ansatz (\ref{msing}) for $m$ into equation (\ref
{eqh}). Since $p>7$ we find from (\ref{msing}) that near zero $%
m_{0}(x)+x^{p}m_{1}(x)=\lambda +\lambda _{0}x^{2}+\lambda _{1}x^{4}+\dots $
. We omit the details of the proof that the asymptotic behavior is as
indicated.

\subsubsection{Existence of the function $\protect\mu_{2}$}

We now prove the existence of a function $m$ that satisfies equation (\ref
{eqh}) with the boundary conditions (\ref{bcm0}) and (\ref{bcm3}). Since the
second derivative of the solution $m$ at zero is positive, and since $m$
converges to zero at infinity, there must be a first $\xi\in\mathbf{R}_{+}, $
such that $m^{\prime}(\xi)=0.$ The basic idea is now to use this position $%
\xi,$ and the value $\rho$ of $m$ at $\xi$, as parameters in shooting
arguments towards zero and infinity. The first shooting argument will allow
us to define a curve $c_{0}$ of initial conditions $(\xi,\rho)$ in $\mathbf{R%
}_{+}^{2},$ for which the boundary condition at zero is satisfied, and the
second shooting argument will allow us to find on this curve an initial
condition for which the boundary condition at infinity is satisfied as well.

So, let $(\xi ,\rho )$ be an initial condition. Locally, i.e., near $\xi ,$
there exists a solution $m_{\xi ,\rho }$ of equation (\ref{eqh}). By
definition, $m_{\xi ,\rho }(\xi )=\rho ,$ $m_{\xi ,\rho }^{\prime }(\xi )=0,$
and therefore we get for the second derivative of $m_{\xi ,\rho }$ at $\xi ,$
\begin{equation*}
m_{\xi ,\rho }^{\prime \prime }(\xi )=\omega _{1}(\xi )\rho ^{n}+\omega
_{2}(\xi )\rho ~,
\end{equation*}
where 
\begin{equation}
\omega _{1}(\xi )=\frac{\left( 2\frac{\mu _{1}(\xi )}{\xi }\right) ^{n}}{\xi
^{2}}~,  \label{omega1}
\end{equation}
and 
\begin{equation}
\omega _{2}(\xi )=\frac{n}{2}\varepsilon -\frac{\delta (\delta +1)}{\xi ^{2}}%
~.  \label{omega2}
\end{equation}
For initial conditions such that $\rho =c_{2}(\xi ),$ where 
\begin{equation}
c_{2}(\xi )=\left( \frac{\frac{n}{2}\varepsilon }{\left( 2\frac{\mu _{1}(\xi
)}{\xi }\right) ^{n}}\left( \xi _{0}^{2}-\xi ^{2}\right) \right)
^{\varepsilon }  \label{c2}
\end{equation}
and $\xi _{0}=\sqrt{\delta (\delta +1)/\left( \frac{n\varepsilon }{2}\right) 
},$ we therefore have that $m_{\xi ,\rho }^{\prime \prime }(\xi )=0.$ See
Fig. 1 for the graph of the function $c_{2}.$ The function $c_{2}$ has a
maximum at the point $\xi _{m}$ that satisfies the equation 
\begin{equation}
\omega _{1}^{\prime }(\xi _{m})c_{2}(\xi _{m})^{n-1}+\omega _{2}^{\prime
}(\xi _{m})=0~,  \label{defxim}
\end{equation}
and the line $c_{2}$ divides the set of initial conditions into two subsets,
a subset $A$ where $m_{\xi ,\rho }^{\prime \prime }(\xi )<0,$ and a subset $%
B $ where $m_{\xi ,\rho }^{\prime \prime }(\xi )>0.$ For initial conditions
on $c_{2}$ we can compute $m_{\xi ,c_{2}(\xi )}^{\prime \prime \prime }(\xi
),$%
\begin{equation*}
m_{\xi ,c_{2}(\xi )}^{\prime \prime \prime }(\xi )=\omega _{1}^{\prime }(\xi
)c_{2}(\xi )^{n}+\omega _{2}^{\prime }(\xi )c_{2}(\xi )~.
\end{equation*}
Comparing with (\ref{defxim}) we find that $m_{\xi _{m},c_{2}(\xi
_{m})}^{\prime \prime \prime }(\xi _{m})=0,$ and we have that $m_{\xi
,c_{2}(\xi )}^{\prime \prime \prime }(\xi )<0$ for $0<\xi <\xi _{m}.$ We now
construct the line $c_{0}$ for $0<\xi <\xi _{m}.$

\begin{proposition}
Fix $\xi,$ $0<\xi<\xi_{m}.$ Then, there exists a unique number $c_{0}(\xi),$ 
$c_{2}(\xi)>$ $c_{0}(\xi)>\lambda,$ such that $m_{\xi,c_{0}(\xi)}$ is
positive and satisfies $\lim_{x\rightarrow0}m_{\xi,c_{0}(\xi)}(x)=\lambda.$
Furthermore, the function $c_{0}$ is continuous.
\end{proposition}

\begin{proof}
The proof is similar to the one in Section \ref{seta777}. Define the two
subsets $I_{1}$ and $I_{2}$ of the interval $I=(\lambda ,c_{2}(\xi ))$, 
\begin{align*}
I_{1}& =\{\rho \in I|~\exists ~0<\xi _{1}<\xi ,~m_{\xi ,\rho }(\xi
_{1})=\lambda \text{ and }\lambda <m_{\xi ,\rho }(x)<c_{2}(x)\text{ for }%
x\in (\xi _{1},\xi )\}~, \\
I_{2}& =\{\rho \in I|~\exists ~0<\xi _{2}<\xi ,~m_{\xi ,\rho }(\xi
_{2})=c_{2}(\xi _{2})\text{ and }\lambda <m_{\xi ,\rho }(x)<c_{2}(x)\text{
for }x\in (\xi _{2},\xi )\}~.
\end{align*}
The intersection of $I_{1}$ with $I_{2}$ is by definition empty, and the
sets $I_{1}$ and $I_{2}$ are open, by continuity of the solution $m_{\xi
,\rho }$ as a function of the initial data $\rho $. We now show that all $%
\rho $ sufficiently close to $\lambda $ are in $I_{1},$ and that all $\rho $
sufficiently close to $c_{2}(\xi )$ are in $I_{2.}$ This implies that $%
c_{0}(\xi )=\sup I_{1}<c_{2}(\xi ),$ and $c_{0}(\xi )$ is neither in $I_{1}$
nor in $I_{2}$, and therefore the function $m_{\xi ,c_{0}(\xi )}$ satisfies $%
\lambda <m_{0}(x)<m_{2}(x)$ for all $0<x<\xi ,$ and therefore $%
\lim_{x\rightarrow 0}m_{\xi ,c_{0}(\xi )}(x)=\lambda ,$ since $%
\lim_{x\rightarrow 0}c_{2}(x)=\lambda .$ So let $(\xi ,\rho )$ be an initial
condition. Then, $m_{\xi ,\rho }$ satisfies the integral equation 
\begin{equation}
m_{\xi ,\rho }(x)=\rho +\int_{\xi }^{x}\frac{dy}{p(y)}\int_{\xi
}^{y}p(z)~\Omega (m_{\xi ,\rho }(z),z)~dz~,  \label{inteq}
\end{equation}
where 
\begin{equation*}
p(z)=\frac{\exp (z^{2}/4)}{z^{2\delta }}~,
\end{equation*}
and where 
\begin{equation*}
\Omega (s,z)=\omega _{1}(z)s^{n}+\omega _{2}(z)s~.
\end{equation*}
$\Omega (s,z)$ is strictly negative for $0<z<\xi _{m}$ and $s\approx \lambda
,$ and therefore we find, like in the proof in Section \ref{theoremeta} that
any solution with an initial condition $\rho $ sufficiently close to $%
\lambda $ will cross the line $m\equiv \lambda .$ Similarly, for an initial
condition $(\xi ,\rho )$ close to $(\xi ,c_{2}(\xi ))$ we can use that $%
\Omega (s,z)\approx $ $0,$ and that $\partial _{z}\Omega (c_{2}(z),z)$ is
strictly negative to show that the corresponding solution will cross the
line $c_{2}.$ This completes the proof of the existence of $c_{0}(\xi ).$ To
prove uniqueness it is sufficient to use that $\partial _{s}\Omega (s,z)>0$
for $(s,z)$ in the set $C$ (see Fig. 1), and to integrate the difference of
two solutions from their respective initial condition to zero, which leads
to a contradiction, since both solutions have to be equal to $\lambda $ at
zero. Finally, that $c_{0}$ is a continuous function follows from the
continuity of $m_{\xi ,\rho }$ as a function of $\rho $ and $\xi $ using the
uniqueness of $c_{0}(\xi ).$
\end{proof}

We now prove with a second shooting argument that solutions with initial
conditions $(\xi,c_{0}(\xi)),$ with $\xi\approx0,$ become negative somewhere
in the interval $(\xi,2),$ and that solutions with initial conditions $%
(\xi,c_{0}(\xi))$, with $\xi\approx\xi_{m},$ stay positive and diverge to
plus infinity.

\begin{proposition}
There exists a unique initial condition $(\xi^{\ast},c_{0}(\xi^{\ast}))$
such that the corresponding solution $m_{\xi^{\ast},c_{0}(\xi^{\ast})}$ is
positive and satisfies $\lim_{x\rightarrow\infty}m_{\xi^{\ast},c_{0}(\xi^{%
\ast})}(x)=0.$
\end{proposition}

\begin{proof}
Define the two subsets $I_{1}$ and $I_{2}$ of the interval $I=(0,\xi _{m})$, 
\begin{align*}
I_{1}& =\{\xi \in I|~\exists ~\xi _{1}>\xi ,~m_{\xi ,c_{0}(\xi )}(\xi _{1})=0%
\text{ and }m_{\xi ,c_{0}(\xi )}(x)>0,\text{ }m_{\xi ,c_{0}(\xi )}^{\prime
}(x)<0\text{ for }x\in (\xi ,\xi _{1})\}~, \\
I_{2}& =\{\xi \in I|~\exists ~\xi _{2}>\xi ,~m_{\xi ,c_{0}(\xi )}^{\prime
}(\xi _{2})=0\text{ and }m_{\xi ,c_{0}(\xi )}(x)>0,\text{ }m_{\xi ,c_{0}(\xi
)}^{\prime }(x)<0\text{ for }x\in (\xi ,\xi _{2})\}~.
\end{align*}
By definition, the intersection of $I_{1}$ with $I_{2}$ is empty, and the
sets $I_{1}$ and $I_{2}$ are open, by continuity of the solution $m_{\xi
,c_{0}(\xi )}$ as a function of the initial data $\xi $. We now show that
all $\xi $ sufficiently close to $0$ are in $I_{1},$ and that all $\xi $
sufficiently close to $\xi _{m}$ are in $I_{2.}$ This implies that $\xi
^{\ast }=\sup I_{1}<\xi _{m},$ is neither in $I_{1}$ nor in $I_{2}$, and
therefore the function $m_{\xi ^{\ast },c_{0}(\xi \ast )}$ is positive and
decreasing for $x>\xi ^{\ast }$ which implies that $\lim_{x\rightarrow
\infty }m_{\xi ^{\ast },c_{0}(\xi ^{\ast })}(x)=0.$ So let $(\xi ,c_{0}(\xi
))$ be an initial condition with $0<\xi <x_{0},$ with $x_{0}\ll 1.$ The
proof that such an initial condition is in $I_{1}$ is rather lengthy and we
therefore do not give the details here, but on a heuristic level it is easy
to understand why such a solution is in $I_{1}$. Namely, near zero the
asymptotics of the solution $m_{\xi ,c_{0}(\xi )}$ is $m_{\xi ,c_{0}(\xi
)}(x)=\lambda +\lambda _{0}x^{2}+\dots +\left( m_{1}\right) _{\xi ,c_{0}(\xi
)}(0)x^{p}+\dots ,$ where $p=p_{+}+\delta ,$ and where $\left( m_{1}\right)
_{\xi ,c_{0}(\xi )}(0)$ is such that $m_{\xi ,c_{0}(\xi )}^{\prime }(\xi )=0$%
. Neglecting higher order terms we find that $m_{\xi ,c_{0}(\xi )}^{\prime
}(\xi )\approx 2\lambda _{0}\xi +\left( m_{1}\right) _{\xi ,c_{0}(\xi
)}(0)p\xi ^{p-1}$, and we conclude that $\left( m_{1}\right) _{\xi
,c_{0}(\xi )}(0)\approx -\left( 2\lambda _{0}/p\right) /\xi ^{p-2}.$
Therefore, $m_{\xi ,c_{0}(\xi )}^{\prime }(\xi ^{(p-5/2)/(p-1)})\approx
-2\lambda _{0}/\xi ^{1/2}\ll 0$ and $m_{\xi ,c_{0}(\xi )}(\xi
^{(p-5/2)/(p-1)})\approx \lambda ,$ if $\xi $ is small enough. Therefore,
since $m_{\xi ,c_{0}(\xi )}^{\prime \prime }(x)<0$ for all $0<x\ll 1,$ we
find that $m_{\xi ,c_{0}(\xi )}(x)<c_{0}(\xi )-2\lambda _{0}(x-\xi )/\xi
^{1/2},$ and therefore $m_{\xi ,c_{0}(\xi )}(x)=0$ for some $x\leq c_{0}(\xi
)\xi ^{1/2}/\left( 2\lambda _{0}\right) +\xi ,$ as claimed. Next, let $(\xi
,c_{0}(\xi ))$ be an initial condition with $\xi _{m}-x_{1}<\xi <\xi _{m},$
with $0<x_{1}<1$ to be chosen below. We now show that such an initial
condition is in $I_{2}.$ For all $\xi _{0}\geq \xi ^{\prime }\geq \xi $ we
have the lower bounds $m_{\xi ,c_{0}(\xi )}(\xi ^{\prime })\geq m_{\xi
,c_{0}(\xi )}(\xi )k_{1}$ and $p(\xi ^{\prime })m_{\xi ,c_{0}(\xi )}^{\prime
}(\xi ^{\prime })\geq $ $m_{\xi ,c_{0}(\xi )}(\xi )k_{2},$ where 
\begin{equation*}
k_{1}=1+\delta (\delta +1)\frac{\exp (\xi _{0}^{2}/4)}{\exp (\xi ^{2}/4)}%
\int_{\xi _{m}-x_{1}}^{\xi _{0}}y^{2\delta }~dy\int_{\xi
_{m}-x_{1}}^{y}z^{-2\delta }~(\frac{1}{\xi _{0}^{2}}-\frac{1}{z^{2}})~dz~,
\end{equation*}
$k_{1}>0,$ and 
\begin{equation*}
k_{2}=\delta (\delta +1)\exp (-\xi _{0}^{2}/4)\int_{\xi _{m}-x_{1}}^{\xi
_{0}}z^{-2\delta }~(\frac{1}{\xi _{0}^{2}}-\frac{1}{z^{2}})~dz~,
\end{equation*}
and for $x\geq \xi _{0}$ we therefore have the lower bound 
\begin{equation*}
m_{\xi ,c_{0}(\xi )}(x)\geq m_{\xi ,c_{0}(\xi )}(\xi )\left(
k_{1}+k_{2}\int_{\xi _{0}}^{\infty }\frac{dy}{p(y)}\right) ~,
\end{equation*}
and it follows, using again the integral equation (\ref{inteq}), that $%
m_{\xi ,c_{0}(\xi )}$ diverges at (or before) infinity, that therefore $%
m_{\xi ,c_{0}(\xi )}^{\prime }(x)=0$ for some $x>\xi ,$ which implies that $%
\xi \in I_{2},$ provided 
\begin{equation}
k_{1}+k_{2}\int_{\xi _{0}}^{\infty }\frac{dy}{p(y)}>0~.  \label{cond}
\end{equation}
For $x_{1}$ small enough and for $n$ large enough (\ref{cond}) can be
verified without too much difficulty. With the help of a computer one can
show that (\ref{cond}) is satisfied for the remaining $n\geq 5.$ For $n=4$ (%
\ref{cond}) is not satisfied, since the above bounds on $m_{\xi ,c_{0}(\xi
)}(\xi _{0})$ and $m_{\xi ,c_{0}(\xi )}^{\prime }(\xi _{0})$ are too weak.
Sufficiently good bounds can be obtained by dividing the interval $(\xi ,\xi
_{0})$ in two pieces and by integrating lower bounds on each of the
subintervals. We omit the details. Finally, uniqueness of $\xi ^{\ast }$ can
be proved by integrating the difference of two solutions from $\xi _{0}$ to
infinity, which, using the positivity of $\partial _{s}\Omega (s,z),$ leads
to a contradiction with the fact that both of the solutions converge to zero
at infinity.
\end{proof}

\subsection{Proof of Proposition \ref{cfi22}}

We first proof the existence of a unique solution of equation (\ref{h}) with
the boundary conditions (\ref{bh-1}) and (\ref{bh-2}). Then, we derive the
results on the asymptotic behavior near zero and infinity.

\subsubsection{Existence of the function $\protect\varphi_{2}$}

The equation (\ref{h}) for $h$ is linear. We therefore first construct two
linearly independent solutions $h_{1}$ and $h_{2}$ for the corresponding
homogeneous equation, which we then use to construct, using standard
methods, a solution of (\ref{h}) that satisfies the boundary conditions (\ref
{bh-1}) and (\ref{bh-2}). The homogeneous equation is 
\begin{equation}
h^{\prime \prime }-q~h=0~,  \label{homh}
\end{equation}
where 
\begin{equation}
q(z)=n\left( 2\kappa z\eta (z)+\eta (z)^{2}\right) ^{n-1}(2\kappa z+2\eta
(z))~.  \label{defq}
\end{equation}
Since the equation (\ref{homh}) is linear, the integral equation for $h_{1}$%
, 
\begin{equation}
h_{1}(x)=1+\int_{0}^{x}~dy\int_{0}^{y}q(z)~h_{1}(z)~dz~,  \label{eqh1}
\end{equation}
has a positive solution that exists for all $x$ in $\mathbf{R}_{+}.$ By
definition, we have near $x=0$ the behavior $h_{1}(x)=1+\mathcal{O}(x^{2}).$
At infinity, the solution $h_{1}$ is asymptotic to a solution of the
equation 
\begin{equation*}
h^{\prime \prime }(x)-\frac{n}{\lambda }(2\kappa \lambda )^{n}\frac{1}{x^{2}}%
h(x)=0~.
\end{equation*}
This equation is the same as the homogeneous part of equation (\ref{eqs}),
and the leading order behavior of $h_{1}$ at infinity is therefore either
proportional to $x^{p_{+}}$ or to $x^{p_{-}},$ with $p_{\pm }$ as defined in
(\ref{eqn:critexp}). Since $h_{1}$ is positive, we find using (\ref{eqh1}),
that $h_{1}(x)>1$ for all $x$ in $\mathbf{R}_{+},$ and therefore $h_{1}$ is
near infinity of the form $h_{1}(x)=d_{1}x^{p_{+}}+\dots ,$ for some
constant $d_{1}>0$. A second solution of the homogeneous equation (\ref{homh}%
) is 
\begin{equation*}
h_{2}(x)=h_{1}(x)~\int_{0}^{x}\frac{1}{h_{1}(y)^{2}}~dy~.
\end{equation*}
Near $x=0$ we have that $h_{2}(x)=x+\dots ,$ and near infinity we find that 
\begin{equation}
h_{2}(x)=h_{1}(x)\left( d-d_{2}x^{1-2p_{+}}+\dots \right) ~,  \label{h2infty}
\end{equation}
where $d=\int_{0}^{\infty }1/h_{1}(y)^{2}~dy$, and $d_{2}=$ $\left(
1/d_{1}\right) ^{2}/\left( 2p_{+}-1\right) .$ We note that $%
h_{1}h_{2}^{\prime }-h_{1}^{\prime }h_{2}\equiv 1.$ Therefore, the function $%
h_{p}$, 
\begin{equation*}
h_{p}(x)=c_{1}(x)~h_{1}(x)+c_{2}(x)~h_{2}(x)~,
\end{equation*}
where 
\begin{align*}
c_{1}(x)& =-\int_{0}^{x}h_{2}(y)~f(y)dy~, \\
c_{2}(x)& =\int_{0}^{x}h_{1}(y)~f(y)dy~,
\end{align*}
and where 
\begin{equation*}
f(x)=-\gamma \eta (x)-\alpha x\eta ^{\prime }(x)+n\left( 2\kappa x\eta
(x)+\eta (x)^{2}\right) ^{n-1}2\kappa _{3}x^{3}\eta (x)~,
\end{equation*}
satisfies equation (\ref{h}). Near zero, the function $f$ is of the form $%
f(x)=-\gamma \eta _{0}+\dots $, and therefore, using the behavior of $h_{1}$
and $h_{2}$ near zero, we find that $c_{1}$ is near zero of order $\mathcal{O%
}(x^{2})$, and $c_{2}$ is near zero of order $\mathcal{O}(x).$ The function $%
h_{p}$ is therefore of order $\mathcal{O}(x^{2})$ near zero. At infinity,
the function $f$ is of the form $f(x)=f_{\infty }x^{-\delta }+\dots ,$ where 
$f_{\infty }=-\gamma \lambda +\alpha \lambda \delta +n(2\kappa \lambda
)^{n-1}2\kappa _{3}\lambda ,$ and therefore the function $c_{2}$ is near
infinity of the form $c_{2}(x)=d_{1}f_{\infty }~x^{p_{+}+1-\delta
}/(p_{+}+1-\delta )+\dots $, and $c_{1}$ is near infinity of the form $%
c_{1}(x)=-d~c_{2}(x)+h_{\infty }+d_{1}d_{2}f_{\infty }~x^{2-p_{+}-\delta
}/(2-p_{+}-\delta )+\dots $, for some constant $h_{\infty }.$ Using these
asymptotic behavior for $c_{1}$, $h_{1}$, $c_{2}$, and $h_{2},$ we find for
the function $h_{p}$ near infinity the behavior, 
\begin{align}
h_{p}(x)& =(-d~c_{2}(x)+h_{\infty }+\frac{d_{1}d_{2}f_{\infty }}{%
(2-p_{+}-\delta )}x^{2-p_{+}-\delta }+\dots )h_{1}(x)+c_{2}(x)h_{1}(x)\left(
d-d_{2}x^{1-2p_{+}}+\dots \right)  \notag \\
& =h_{\infty }h_{1}(x)-\frac{d_{1}f_{\infty }}{p_{+}+1-\delta }%
d_{1}d_{2}x^{p_{+}+1-\delta }x^{p_{+}}x^{1-2p_{+}}+\frac{d_{1}d_{2}f_{\infty
}}{2-p_{+}-\delta }d_{1}x^{2-p_{+}-\delta }x^{p_{+}}+\dots  \notag \\
& =h_{\infty }h_{1}(x)+\frac{f_{\infty }}{2p_{+}-1}\left( \frac{-1}{%
p_{+}+1-\delta }+\frac{1}{2-p_{+}-\delta }\right) x^{2-\delta }+\dots  \notag
\\
& =h_{\infty }h_{1}(x)+\lambda _{0}x^{2-\delta }+\dots ~.  \label{hpinfty}
\end{align}
In the last equality we have used the definition (\ref{lambda0}) for $%
\lambda _{0}.$ The function $h,$%
\begin{equation*}
h(z)=h_{p}(z)-h_{\infty }~h_{1}(z)~,
\end{equation*}
solves the equation (\ref{h}), satisfies the boundary condition (\ref{bh-1}%
), and since, as we show in the next section, the higher order terms in (\ref
{hpinfty}) converge to zero at infinity, it also satisfies the boundary
condition (\ref{bh-2}).

\subsubsection{Asymptotic behavior of the function $\protect\varphi_{2}$}

By construction, the leading behavior of $h$ at infinity is $h(z)=\lambda
_{0}z^{2-\delta }+\dots $. We therefore make the ansatz $h(z)=\lambda
_{0}z^{2-\delta }+k(z)$, and to leading order, we get for the function $k$
the linear equation 
\begin{equation}
k^{\prime \prime }-n\delta (\delta +1)z^{-2}k=c_{k}z^{-\delta ^{\prime }}~,
\label{bvarphi2-2}
\end{equation}
for a certain constant $c_{k}$. The general solution of equation (\ref
{bvarphi2-2}) is 
\begin{equation*}
k(z)=\frac{\lambda ^{\prime }}{z^{\delta ^{\prime }-2}}+\mathrm{const.}%
~z^{p_{-}}+\mathrm{const.}~z^{p_{+}}~,
\end{equation*}
with a certain constant $\lambda ^{\prime }$ and with $p_{+}$, $p_{-}$ as
defined in (\ref{eqn:critexp}). Since $\lim_{z\rightarrow \infty
}k(z)/z^{2-\delta }=0$ but $p_{+}>2-\delta ,$ the coefficient of the term
proportional to $z^{p_{+}}$ must be zero. Therefore, since$\left|
p_{-}\right| >\delta ^{\prime }-2$ for all $n,$ the asymptotic behavior of $%
k $ is always given by $\lambda ^{\prime }/z^{\delta ^{\prime }-2}.$ We omit
the details of the proof that this is indeed the correct second order
behavior of $k$ at infinity.\bigskip \bigskip \bigskip

\noindent \textbf{Acknowledgment}

During this work, A.S. was hosted by the University of Chicago and wishes to
thank the Department of Mathematics for its hospitality.

\end{document}